\documentclass[letterpaper]{JHEP3}
\usepackage{graphicx}
\newcommand{\beq}{\begin{equation}}
\newcommand{\eeq}{\end{equation}}

\usepackage{epsfig,multicol}

\title{Twisted supersymmetric sigma model on the lattice}
\author{Simon Catterall\\ Department of Physics, Syracuse University, Syracuse, NY13244, USA\\
E-mail: smc@phy.syr.edu}
\author{Sofiane Ghadab\\ Department of Physics, Syracuse University, Syracuse, NY13244, USA\\
E-mail: sghadab@phy.syr.edu}
\abstract{In this paper we conduct a numerical study of the
supersymmetric $O(3)$ non-linear sigma model. The lattice
formulation we employ was derived in \cite{sigma1} and corresponds
to a discretization of a {\it twisted} form of the continuum action. 
The twisting process exposes a {\it nilpotent} supercharge $Q$ and allows
the action to be rewritten in $Q$-exact form. These properties may be
maintained on the lattice. We show how to deform the theory
by the addition of potential terms which preserve the supersymmetry.
A Wilson mass operator may be introduced in this way with
a minimal breaking of supersymmetry.
We additionally show how to rewrite the theory
in the language of K\"{a}hler-Dirac fields and explain why this
avenue does not provide a good route to discretization.
Our numerical results provide strong evidence 
for a restoration of full supersymmetry
in the continuum limit {\it without} fine tuning.
We also observe a non-vanishing chiral condensate as expected from
continuum instanton calculations.}
\preprint{SU-4252-833}

\begin{document}
\section{Introduction}

Supersymmetric field theories are an integral part of modern theories of particle 
physics. They provide a framework for solving the gauge hierarchy problem 
\cite{hierarchy} by eliminating many divergences typical of quantum field theories
through the cancellation between fermionic and bosonic loops. Moreover in the 
large $N$ limit, it is known \cite{large_N,Maldacena} that supersymmetric gauge 
theories are related to quantum gravity and string theory. Two dimensional sigma models
on the other hand are important because they have a rich mathematical structure, 
and moreover there exists a deep rooted analogy 
between them and four dimensional Yang Mills theories \cite{polyakov}. Indeed the 
former can serve as perfect theoretical laboratories to test methods and approaches
developed for solving the problems of these far richer and more complicated theories.
For example, the magnetic monopole and dyon in the gauge theory correspond to the kink
and Q-kink solution of sigma models. In many cases the interesting physics
lies in non-perturbative regimes which motivates use of the lattice to study
these systems.       

Unfortunately generic discretizations of supersymmetric
theories break supersymmetry explicitly and necessitate fine
tuning the couplings to a (usually large) number
of lattice operators in order to approach a supersymmetric
continuum limit. Recently, attention has turned to formulations 
which aim to preserve one or more exact
supersymmetries at non-zero lattice spacing\footnote{We primarily discuss
Euclidean formulations - for recent work on Hamiltonian approaches see
\cite{Harada}}. The hope is, this
residual supersymmetry will shield us 
from the appearance of relevant operators in 
the lattice effective action which violate the 
full symmetry group thus
reducing or possibly even eliminating such fine tuning problems.
Two distinct approaches in this direction have been followed; in the first 
pioneered by Kaplan and collaborators, the lattice theory is constructed by orbifolding
a certain supersymmetric matrix model and then extracting the lattice theory by expansion 
around some vacuum state \cite{orbifold1,orbifold2,orbifold3,orbifold4}. This 
has recently been extended to the the interesting case of gauge theories in two dimensions with
matter fields interacting via a superpotential in \cite{orbifold5}. The second approach 
relies on discretizing a \emph{twisted} version of the supersymmetric theory. 
\emph{Twisting} was first introduced by Witten \cite{witten}
in the context of topological field theories. It consists of 
constructing a new rotation group from a combination of the original 
rotation group and part of the $R$-symmetry associated with the extended SUSY. 
The 
supersymmetric field theory is then reformulated in terms of fields which 
transform as 
integer representations of this new rotation group 
\cite{Simon03,Kato03,D'Adda04,Kato05}. In flat space 
one can think of the twisting as a merely exotic change of variables in the theory.
When applied to the supersymmetry algebra, a scalar nilpotent supercharge is exposed.
Furthermore as argued in \cite{Simon04_1,Simon04_2} the twisted superalgebra implies 
that the action rewritten in terms of these twisted fields is generically $Q$-exact. In 
this case it is straightforward to construct a lattice action which is $Q$-invariant 
provided \emph{only} that we preserve the nilpotency of $Q$ under discretization. 
An other approach in the same direction \cite{D'Adda04,Kawamoto}, 
attempts to preserve all the twisted
supercharges on the lattice by introducing a 
non-trivial commutation relation between the coordinates in
superspace. 
However, this formulation remains controversial after a recent paper 
\cite{Bruckmann06} pointed out an inconsistency appearing in this method
when applied to a toy quantum mechanical model.
Finally, although the orbifolding and the twisting approaches appear different, 
\"Unsal \cite{Unsal06} recently showed that the former approach
reproduces the twisted
Yang-Mills theories in the continuum limit.  

The \emph{twisting} approach to constructing lattice theories
was initially 
developed for theories without gauge symmetry
\cite{Simon00,Simon01}. An implementation for 
supersymmetric lattice gauge theories based on 
balanced topological field theories
was given by Sugino \cite{Sugino}. An approach emphasizing the
geometrical nature of the twist and employing 
K\"ahler-Dirac fermions was then
developed by Catterall 
\cite{Simon04_1, Simon_kahler}
to construct super Yang-Mills theories in two and four dimensions. 
The use of K\"{a}hler-Dirac fermions for formulating lattice supersymmetry
was first proposed in \cite{schwimm}.

\section{The 2D continuum action}
As was shown in \cite{sigma1}, 
the action of a general two 
dimensional sigma model with ${\cal N}=2$
supersymmetry may be written, using complex
coordinates, in the twisted
form 
\begin{eqnarray}
S&=&\beta\int d^2\sigma\left(
2h^{+-}g_{I\overline{J}}\partial_+\phi^I
\partial_-\phi^{\overline{J}}\right.\nonumber\\
&-&\left.h^{+-}g_{I\overline{J}}\eta^I_+D_-\psi^{\overline{J}}-
h^{+-}g_{\overline{I}J}\eta^{\overline{I}}_-D_+\psi^J+
\frac{1}{2}h^{+-}R_{I\overline{I} J\overline{J}}
\eta^I_+\eta^{\overline{I}}_-\psi^J\psi^{\overline{J}}\right)
\label{2daction}
\end{eqnarray}
with
\begin{equation}
D_+\psi^J=\partial_+\psi^J+\Gamma^J_{KL}\partial_+\phi^K\psi^L
\end{equation}
The requirement of ${\cal N}=2$ supersymmetry forces the target
manifold to be K\"{a}hler in which case
$g_{IJ}=g_{\bar{I}\bar{J}}=0$ and the only non-zero 
Christoffel symbols are $\Gamma^I_{JK}$ and $\Gamma^{\bar{I}}_{\bar{J}\bar{K}}$ (see
Appendix \ref{nota}). This
action is invariant under four supersymmetries as expected for
a theory with ${\cal N}=2$ supersymmetry in two
dimensions. In the twisted construction we focus on a single
hermitian twisted supercharge $Q$ whose action on the fields is
given by
\begin{equation}
\left\{
\begin{array}{lll}
Q\phi^I = \psi^I &\quad& Q\phi^{\bar{I}} = \psi^{\bar{I}}\\
Q\psi^I = 0 &\quad& Q\psi^{\bar{I}}= 0\\
Q\eta^I_+=B^I_+-\Gamma^I_{KJ}\psi^J\eta^K_+ &\quad& 
    Q\eta^{\bar{I}}_-=B^{\bar{I}}_--\Gamma^{\bar{I}}_{\bar{K}\bar{J}}
\psi^{\bar{J}}\eta^{\bar{K}}_-\\
QB^I_+ = -\Gamma^I_{JK}\psi^JB^K_+-R^I_{\;\;K\bar{J}L}\psi^K\psi^{\bar{J}}\eta^L_+ &\quad& 
    QB^{\bar{I}}_- = -\Gamma^{\bar{I}}_{\bar{J}\bar{K}}\psi^{\bar{J}}
    B^{\bar{K}}_--R^{\bar{I}}_{\;\;\bar{J}L\bar{K}}\psi^{\bar{J}}\psi^L
\eta^{\bar{K}}_-\\
\end{array}
\label{Qtrans}
\right.
\end{equation}
The action of the remaining charges of the $N=2$ twisted supersymmetry is given in appendix 
\ref{twisted}. The field $B^I_+$ is an auxiliary field
introduced to linearize the transformations and render the 
transformation nilpotent off-shell. It has been removed from
the action in eqn.~(\ref{2daction}) by employing its equation of motion
$B^I_+=\partial_+\phi^I$.
The invariance of the action under this 
supercharge $Q$ follows 
just from the nilpotency of the latter and the fact that the 
above action is $Q$-exact -- that is it can be written as the
$Q$-variation of some function which, borrowing from BRST gauge fixing
terminology, can be termed a gauge fixing
fermion \cite{sigma1}.

\section{The 2D lattice action}
Surprisingly translating the action in (\ref{2daction}) to the lattice is pretty
straightforward. Indeed, as the twisted
$Q$-symmetry makes no reference to derivatives 
of the fields its
nilpotent property is preserved when continuum fields
indexed by a continuous coordinate are replaced by
lattice fields carrying a discrete index.
We then obtain the supersymmetric
lattice action by just replacing the continuum derivative 
by a symmetric finite difference 
\begin{eqnarray}
S&=&\beta\sum_x\left(
2h^{+-}g_{I\overline{J}}\Delta_+^s\phi^I
\Delta_-^s\phi^{\overline{J}}\right.\nonumber\\
&-&\left.h^{+-}g_{I\overline{J}}\eta^I_+D_-\psi^{\overline{J}}-
h^{+-}g_{\overline{I}J}\eta^{\overline{I}}_-D_+\psi^J+
\frac{1}{2}h^{+-}R_{I\overline{I} J\overline{J}}
\eta^I_+\eta^{\overline{I}}_-\psi^J\psi^{\overline{J}}\right)
\label{2dLaction}
\end{eqnarray} 
where now,
\begin{equation}
D_+\psi^J=\Delta_+^s\psi^J+\Gamma^J_{KL}\Delta_+^s\phi^K\psi^L
\label{Deriv}
\end{equation}
and $\Delta^s_\pm=\Delta^s_1\pm i\Delta^s_2$ where $\Delta^s_\mu=\frac{1}{2}
(\Delta^+_\mu+\Delta^-_\mu)$ and $\Delta^{\pm}$ are the usual
forward and backward difference operators.  
However the lattice action in the form shown in eqn.~(\ref{2dLaction})  suffers from the usual fermion doubling 
problem. Indeed it is easy to see that the kernel of the (free) lattice Dirac operator constructed 
this way contains extra states which have no continuum interpretation. 
For instance consider the fermionic part in (\ref{2dLaction}). If we ignore the interaction term 
in (\ref{Deriv}) then one finds that in Fourier space the kernel of $\Delta^s_+$ corresponds to solving
\begin{equation}
0 = e^{ikx}[\sin k_1+i \sin k_2]
\end{equation} 
Besides the solution $(k_1,k_2)=(0,0)$, one also has $(0,\pi),(\pi,0)$ and $(\pi,\pi)$.
The doubling problem is 
made worse because supersymmetry propagates it to the bosonic sector. 
There are two obvious lines of approach one may take to avoid
this problem. One is rewrite the continuum theory in the language of
K\"{a}hler-Dirac fields and try to utilize
the discretization procedure developed in \cite{Simon04_1}
to construct the lattice theory. This approach is reviewed
in the next section where it is shown that such a procedure runs into
difficulties. 

The obvious alternative, which we have
pursued here, is to add 
some form of Wilson mass term to lift the mass of the doubles
up to the scale of the cut-off. However an ad hoc addition of
such a Wilson term will 
break supersymmetry explicitly in an uncontrolled way
thus spoiling our goal of 
preserving an element of 
SUSY on the lattice. Fortunately, 
it was shown \cite{labast, labast2} that in the case where the 
K\"{a}hler manifold possesses some isometries, 
it is possible to add potential terms that involve 
the holomorphic Killing vectors $V^I$ associated to those isometries in such
a way as to keep
the action $Q$-exact. 
By a careful choice of such potential terms we may
add Wilson operators which accomplish the task of
rendering the doublers heavy.
In complex coordinates the 
possible terms are,
\begin{equation}
\Delta S= \beta \sum_x \left[ \lambda^2
V^IV_I+\lambda^2\psi^{\overline{I}}\nabla_{\overline{I}}V_{J}\psi^{J}-
\frac{1}{4}h^{+-}\eta_-^{\overline{I}}\nabla_{\overline{I}}V_{J}\eta_+^{J}+{\rm h.c} \right]
\label{killing}
\end{equation}
Here, $\lambda^2$ is an arbitrary parameter. Here, we choose $\lambda=h^{+-}=\frac{1}{2}$.
To keep the action $Q$-exact the action of the twisted supersymmetry
must be changed.
\begin{equation}
\left\{
\begin{array}{l}
Q\phi^I = \psi^I \\
Q\psi^I = V^I \\
Q\eta^I_+=B^I_+-\Gamma^I_{JK}\psi^J\eta^K_+ \\
QB^I_+ = -\Gamma^I_{JK}\psi^JB^K_++\frac{1}{2}R^I_{\;\;JKL}\psi^K\psi^{J}
\eta^L_++D_KV^I\eta^K_+ 
\end{array}
\right.
\label{deform}
\end{equation}  
$Q$ 
is no longer nilpotent but $Q^2$ just amounts to a Lie 
derivative with respect to the 
Killing vector field.
Indeed one can easily show that,
\begin{equation}
\left\{
\begin{array}{lll}
Q^2\phi^I &=& V^I \\
Q^2\psi^I &=& \partial_JV^I \psi^J\\
Q^2\eta^I_+ &=& \partial_JV^I\eta^J_+ \\
Q^2B^I_+ &=& \partial_J V^I B^J_+ 
\end{array}
\right.
\end{equation}
To introduce a mass term would require utilizing a Killing vector of the form
\beq V^I=im\phi^I\label{kill}\eeq
It should be
clear that the transformation induced by such a Killing vector corresponds
to a infinitessimal phase rotation on the complex
fields and can be seen to be a good symmetry of both
the continuum {\it and} lattice actions.
Many K\"ahler manifolds possess metrics invariant under such
a phase rotation, for example, the $CP^N$ 
theories and specifically the $CP^1\sim O(3)$ model considered in detail later.
In the latter case the theory actually possesses three
isometries corresponding to the global $O(3)$ symmetry of
the theory - the phase rotation associated with the
Killing vector eqn.~\ref{kill} corresponds to invariance under
rotations about the z-axis.

To remove the doubles in lattice regularizations of such models
we have employed such a twisted mass term with the constant
mass parameter replaced by a Wilson
operator $m\to m_W$ where
\begin{equation}
\label{wilson}
m_W=\frac{1}{2}\left(\Delta^+_+\Delta^-_-+
                     \Delta^+_-\Delta^-_+\right)
\end{equation}
This leads to the additional terms in the action
\begin{equation}
\Delta S= \beta \sum_x \left[ \lambda^2
m_W\phi^Im_W\phi_I+\lambda^2\psi^{\overline{I}}\nabla_{\overline{I}}
g_{J\overline{J}} m_W \phi^{\overline{J}}\psi^{J}-
\frac{1}{4}h^{+-}\eta_-^{\overline{I}}\nabla_{\overline{I}}
g_{J\overline{J}} m_W \phi^{\overline{J}}\eta_+^{J} +{\rm h.c}\right]
\end{equation}
The generalized phase rotation associated with
$m_W$ {\it no longer} corresponds to an exact isometry of the metric and
hence the modified action, while still $Q$-exact, 
is no longer exactly invariant under the
generalized $Q$-symmetry. Novertheless we will argue that
the twisted Wilson operator acts as a soft breaking term and
hence should not affect the renormalization of the lattice
theory for small enough lattice spacing.

The argument goes as follows.
In the continuum one is used to thinking of
mass terms as serving only to break supersymmetry {\it softly}.
This expectation 
relies on the idea that any mass parameter will be
small compared to the U.V cut-off
in the theory. However, the generic Wilson operator
in a lattice theory does {\it not} satisfy this property since the potential
doublers in the theory pick up a mass from the Wilson term on the order of the
cut-off. Thus addition of
such Wilson terms generically will lead to a hard breaking of
supersymmetry and it will be necessary to fine tune additional 
operators to recover a supersymmetric continuum limit. 
However, the twisted mass operator we use here does not have this
property -- in the limit of small $a$ the propagators are dominated
by contributions from a state near the origin of the 
Brillioun zone and a set of would be 
doublers. The latter contribution does not break supersymmetry since
the doublers contribute like additional states with a {\it constant} 
twisted mass. Such a mass deformation preserves the
$Q$-symmetry of the lattice action.  

The soft character of the
breaking can be seen in yet another way. For small lattice
spacing $a$ (large $\beta$) we can expand the
bosonic action to quadratic order. Subsequently integrating
over the bosons yields
a determinant ${\rm det}(m_W^2-D^S_+D^S_-)$. But at one loop
this
determinant is cancelled by an
identical fermionic contribution ${\rm det} (M)$ where
\[M=\left(\begin{array}{ll}
m_W&-D^S_+\\
-D^S_{-}&m_W
\end{array}\right)
\]
Thus the lattice theory appears both double free and
supersymmetric at large $\beta$.  
These arguments lead us to expect the $Q$-symmetry to
be restored at large coupling $\beta$ without additional
fine tuning. If this is the case then simple power counting arguments
lead us to conclude that the only (marginally)
{\it relevant} counter terms must take the $Q$-exact form
\begin{equation}
O=Q\left(\eta^{i\alpha}f_{ij}\left(\phi\right)
\left(\partial_\alpha\phi^j+B^j\right)\right)
\end{equation}
where we have reverted to a formulation using real fields.
General covariance ensures that $f_{ij}$  
is a tensor which may then be taken to represent a quantum
renormalization of the target space metric tensor. This
counterterm structure would be consistent with
a lattice model which exhibits ${\cal N}=1$ supersymmetry in the
continuum limit. The restoration of
full ${\cal N}=2$ supersymmetry appears to require additional constraints.
Luckily,
such constraints are present in the form of additional discrete
symmetries
of the lattice action. Consider the classical action
in K\"{a}hler form given in eqn.~\ref{2dLaction}. 
It is trivial to see that this action is also invariant 
under the finite transformations
\begin{eqnarray}
\psi^I&\to & i\psi^I\nonumber\\
\psi^{\overline{I}}&\to & -i\psi^{\overline{I}}\nonumber\\
\eta^I_+&\to & i\eta^I_+\nonumber\\
\eta^{\overline{I}}_-&\to & -i\eta^{\overline{I}}_-
\label{sym}
\end{eqnarray}
Actually, this additional symmetry arises from the K\"{a}hler structure
of the target space appearing in the classical action \cite{alvarez}.
This additional 
symmetry of the lattice model then ensures that only
counterterms compatible with a K\"{a}hler target space survive in the
quantum effective action.
But as was shown in \cite{alvarez} any
model with ${\cal N}=1$ supersymmetry and a K\"{a}hler target space automatically
possesses ${\cal N}=2$ supersymmetry. Thus on the
basis of these arguments we expect that no additional
fine tuning is needed to regain the full supersymmetry of the continuum
model which as will be seen later, is confirmed by our numerical results.

\section{K\"{a}hler-Dirac reformulation}
Recent work has emphasized the
geometrical nature of certain twisted super Yang-Mills theories
\cite{Simon04_1,Simon_kahler}. In these
constructions all fields appear as antisymmetric tensors
with the twisted fermions 
arising as components of a geometrical object called
a K\"{a}hler-Dirac field. The arguments leading to this
geometrical interpretation are quite general, depending only
in the number of supercharges and $R$-symmetry, and imply
that it should be possible
to formulate the twisted sigma models also in this
language.
Furthermore, the geometrical
construction is key to
a successful discretization of the Yang-Mills theory avoiding fermion
doublers \cite{rabin}.
Thus such a reformulation of the sigma model
potentially offers another way to 
build a supersymmetric lattice theory without
encountering fermion doubling. As such it would offer an
alternative to the addition of Wilson operators as described
in the previous section.
As we will see this reformulation
can indeed be done in the continuum but an obstruction prevents
any simple translation of the continuum theory to the lattice.

As shown in, for example, \cite{Simon_kahler} the basic ${\cal N}=2$
twisted supermultiplet in two
dimensions involves two K\"{a}hler-Dirac fields, a
bosonic one $\Phi=(\phi,A_\mu,B_{12})$ whose p-form components commute
and fermionic one $\Psi=(\lambda,\eta_\mu,\chi_{12})$ with
grassmann valued forms. Initially we will consider a flat
target space and consider only the continuum theory.
The action of the nilpotent twisted supersymmetry is simply
\beq
\begin{array}{ll}
Q\phi=\lambda&\quad Q\lambda=0\\
Q\eta_\mu=A_\mu&\quad Q A_\mu=0\\
QB_{12}=0&\quad Q\chi_{12}=0
\end{array}
\eeq
These transformations are essentially the same as the corresponding
Yang-Mills variations except that the field $A_\mu$ plays the
role of a multiplier field in the sigma model case (and hence
has zero variation under $Q$).
The twisted action can be written as the $Q$-variation
of some gauge fermion $\Lambda$. The most general 
gauge fermion which is
linear in the fermion fields, contains at most
one derivative, and is not a $Q$-singlet is
\beq
\Lambda=\int d^2x \eta_\mu\left[c_1A_\mu+c_2\partial_\mu\phi+c_3\partial_\nu B_{\mu\nu}\right]
\eeq
Furthermore, by a simple rescaling of the fields I can set $c_1=\frac{1}{2}$
and $c_2=c_3=1$.
Carrying out the $Q$-variation and integrating out the multiplier field
$A_\mu$ (along the imaginary axis) leads to the action
\beq
S=\int d^2x \left[\frac{1}{2}\left(\partial_\mu\phi+\epsilon_{\mu\nu}\partial_\nu
B_{12}\right)^2-\eta_\mu\partial_\mu\lambda-\chi_{\mu\nu}
\partial_{\left[\mu\right.}\eta_{\left.\nu\right]}\right]
\label{kdaction}
\eeq
As in \cite{Simon_kahler} the twisted fermionic action is of K\"{a}lher-Dirac
form.
Writing $u^1=\phi$, $u^2=B_{12}$ the bosonic action can now be
rewritten
\beq
S_B=\int d^2 x \frac{1}{2}\left[\left(\partial_\mu
u^i\right)^2+\epsilon_{ij}\epsilon_{\mu\nu}\partial_\mu u^i\partial_\nu u^j
\right]
\eeq
where the implied summations over Roman indices run from $i=1,2$.
This action can be simplified by introducing the projection
operator
\beq
P^{+i\nu}_{j\mu}=\frac{1}{2}\left(\delta^i_j\delta^\nu_\mu+
\epsilon^i_j\epsilon^\nu_\mu\right)\eeq
where the distinction between upper and lower indices is, for the
present, immaterial.
The bosonic action now reads
\beq
S_B=\int d^2 x \left[P^{+i\nu}_{j\mu}\partial_\nu u^j\right]^2\eeq
Proceeding in this way we can introduce a 
new grassman valued field which is the superpartner to
$u^i$ by defining $\psi^1=\lambda$ and $\psi^2=\chi_{12}$.
Similarly the original field $A_\mu$ (and its
partner $\eta_\mu$) can be promoted to a field
indexed by a Roman target index $A_\mu\to A^{+i}_\mu$
{\it provided} the field is required to be self-dual under the
action of this projector i.e
\beq
A^i_\mu=A^{+i}_{\mu}=P^{+i\nu}_{j\mu}A^{j}_\nu\eeq
In terms of these new variables the
$Q$-transformations are now
\beq
\begin{array}{ll}
Qu^i=\psi^i&\quad Q\psi^i=0\\
Q\eta^{+i}_\mu=A^{+i}_\mu&\quad QA^{+i}_\mu=0\\
\end{array}
\eeq
Going to complex coordinates we recover our original sigma model 
variations for a flat two dimensional target space.
It should be clear that the sigma model fields are essentially
the
the Hodge self-dual components of the original K\"{a}hler-Dirac fields.
It should also be clear how to proceed to a more general
target space. Clearly one must be able to introduce a tensor
$J^i_j$, here just $\epsilon^i_j$, which squares to minus the identity
and is covariantly
constant (so as not to disturb the form of the $Q$-transformations
when applied to the self-dual fields). These requirements require
the target manifold be K\"{a}hler with $J^i_j$ its complex structure. 
Of course these were just the restrictions found in earlier
constructions of the twisted sigma models \cite{sigma1}.
Keeping the simple form of the $Q$-transformations the twisted
sigma model action can be obtained as
\beq
S=Q\int d^2 x \eta^{+i}_\mu\left[\partial_\mu
u^j-\frac{1}{2}A^{+j}_\mu-\Gamma^j_{lk}\eta^{+l}_\mu\psi^k\right]g_{ij}\eeq

Having recast the continuum theory in this language we can return the
problem of constructing a corresponding lattice model.
We see an immediate problem. To avoid
spectrum doubling we must discretize continuum derivatives with care.
Specifically an exterior derivative must be replaced by a {\it forward}
difference operator $D^+$ in the discrete
theory while the adjoint of an exterior derivative
by a {\it backward} difference operator $D^-$ if
the resulting lattice theory is to be free of spectrum
doubling \cite{rabin}.
Thus the bosonic part of the action in eqn.~(\ref{kdaction}) must be
replaced with
\beq
S_B=\sum \frac{1}{2}\left(D^+\phi+D^-B_{\mu\nu}\right)^2\eeq
Since the difference operators are no longer the same we can no longer
introduce the projector $P^+$ and rewrite the theory in terms of
self-dual fields. Thus the K\"{a}hler-Dirac approach cannot be used
to construct twisted lattice theories without spectrum 
doubling. A similar phenomena is encountered in the Yang-Mills
case where the self-dual nature of the fields in four
dimensional ${\cal N}=2$
super Yang-Mills prevents construction of a $Q$-exact and
double free lattice theory.

\section{Simulations}
Here we revisit the $O(3)$ sigma model discussed in detail in \cite{sigma1}. 
The $O(3)$ sigma model has the 
advantage of being simple and can be reformulated as a twisted model with a $CP^1$ 
target space. The metric, connection and curvature take the form,
\begin{eqnarray}
g_{u\overline{u}}&=&\frac{1}{2\rho^2}\\
\Gamma^u_{uu}&=&g^{\overline{u}u}\partial_u g_{\overline{u}u}=-\frac{2\overline{u}}{\rho}\\
R_{\overline{u}u\overline{u}u}&=&g_{\overline{u}u}\partial_{\overline{u}}\Gamma^u_{uu}
=-\frac{1}{\rho^4}
\end{eqnarray} 
where $\rho=1+u\overline{u}$. From (\ref{2dLaction}), (\ref{killing}) and (\ref{wilson}) 
the (effective) lattice action for a flat base space is given by (here $I=\bar{I}=1$),
\begin{eqnarray}
\label{sigma_action}
S&=&\sum_{x}\left\{ \beta \left[\frac{1}{\rho^2}\Delta^S_+u\Delta^S_-\overline{u}
+\frac{1}{\rho^2}(m_Wu)(m_W\overline{u}) 
-\frac{1}{2\rho^2}\eta D^S_-\overline{\psi}-\frac{1}{2\rho^2}\overline{\eta} D^S_+\psi
-\frac{1}{2\rho^4}\overline{\eta}\eta\overline{\psi}\psi
\right. \right. \nonumber \\
&&
+i\overline{\psi} [\frac{1}{2\rho^2}m_W+m_W \frac{1}{2\rho^2}-\frac{\bar{u}}{\rho^3}(m_Wu)
-\frac{u}{\rho^3}(m_W\bar{u})]\psi \nonumber \\
&& \left. \left. - \frac{i}{4\rho^2}\overline{\eta}
[\frac{1}{2\rho^2}m_W+m_W \frac{1}{2\rho^2}-\frac{\bar{u}}{\rho^3}(m_Wu)
-\frac{u}{\rho^3}(m_W\bar{u})]\eta
\right]+\ln[\beta /(2\rho^2)]\right\} 
\end{eqnarray}
where a factor of two has been absorbed into the coupling $\beta$ and we have simplified our
notation by replacing $\phi^1\rightarrow u$,$\psi^1\to\psi$,
$\eta_+^1\to\eta$ and $\overline{\eta}_-^{\bar{1}}\to\overline{\eta}$. 
The explicit
form of the lattice covariant derivative is 
\begin{equation}
D^S_+=\Delta^S_+ -\frac{2\overline{u}}{\rho}\left(\Delta^S_+ u\right)
\end{equation}
Note that the new term $\ln[\beta /(2\rho^2)]$  missing in \cite{sigma1}, emerges from 
integrating out the auxiliary fields $B^i_+$\footnote{We thank Joel Giedt
and Erich Poppitz for pointing out this factor}.  
To proceed further it is convenient to introduce an other auxiliary field $\sigma$
to remove the quartic fermion term. Explicitly we employ the
identity
\begin{equation}
\beta^V\int D\sigma e^{-\alpha\left(\frac{1}{2}\sigma\overline{\sigma}+
\frac{\sigma}{2\rho^2}\overline{\eta}\psi+
\frac{\overline{\sigma}}{2\rho^2}\eta\overline{\psi}\right)}=
e^{\frac{\alpha}{2\rho^4}
\overline{\eta}\eta\overline{\psi}\psi}
\end{equation}
where $V$ is the number of lattice sites. Thus the partition function
of the lattice model can be cast in the
form
\begin{equation}
Z=\int DuD\sigma D\eta D\psi e^{-S\left(u,\sigma,\eta, \psi\right)}
\end{equation}
where the action is now given by
\begin{eqnarray}
S=\beta\sum_{x}\left[\frac{1}{\rho^2}\Delta^S_+u\Delta^S_-\overline{u}+
\frac{1}{\rho^2}(m_Wu)(m_W\overline{u})
+\frac{1}{2}\sigma\overline{\sigma}-\frac{1}{\beta}\ln (2\rho^2)+
\overline{\Psi} M(u,\sigma) \Psi \right]
\label{o3act}
\end{eqnarray}
where we have assembled the
twisted fields into Dirac spinors 
\[\begin{array}{ccc}
\overline{\Psi}=\left(\begin{array}{c}\frac{\bar{\eta}}{2i}\\
\bar{\psi}\end{array}\right)&\quad&
\Psi=\left(\begin{array}{c}
\frac{\eta}{2i} \\ \psi \end{array}\right)
\end{array}
\]
and the Dirac operator $M(u,\sigma)$ in this chiral basis is
\begin{equation}
M(u,\sigma)=\left(
\begin{array}{cc}
\frac{1}{2\rho^2}m_W-\frac{\bar{u}}{\rho^3}(m_Wu)+{\rm h.c}& \frac{1}{\rho^2}(\Delta^S_+ -\frac{2\overline{u}}{\rho}
\left(\Delta^S_+u\right)+\sigma) \\
(\Delta^S_- +\frac{2u}{\rho}
\left(\Delta^S_-\bar{u}\right)-\bar{\sigma})\frac{1}{\rho^2} 
& \frac{1}{2\rho^2}m_W-\frac{\bar{u}}{\rho^3}(m_Wu)+{\rm h.c}
\end{array}\right)
\end{equation}
Notice the extra terms depending on the target space connection
appearing along the diagonal which correspond to the
twisted Wilson mass operator. These were not present in our previous
paper \cite{sigma1}.

In order to be able to simulate this model, one needs to rewrite the effective action 
in a form that doesn't involve the grassman fields. For that we integrate out the 
Dirac field $\Psi$ generating,
\begin{equation}
\label{fermion_det}
{\rm det}^{\frac{1}{2}}(\beta^2M(u,\sigma)^\dagger M(u,\sigma))
\end{equation}
The effective 
action is now given by,
\begin{equation}
\label{effective}
S=\beta S_B(u,\sigma)-\ln(2\rho^2)-\frac{1}{2}{\rm Tr}
\ln{\left(\beta^2M^\dagger(u,\sigma)M(u,\sigma)\right)}
\end{equation}
Clearly, the form of the fermion effective action we employ does not take into account
any non-trivial phase associated with the fermion determinant - our simulation 
generates the phase quenched ensemble. We later examine the phase explicitly. 

To simulate this model we used the RHMC algorithm developed by Clark and
Kennedy \cite{Clark}. The first step of this algorithm replaces the fermion determinant by an
integration over auxiliary \emph{commuting} pseudofermion fields $F$ and $F^\dagger$ in 
the following way,
\begin{equation}
det^{\frac{1}{2}}(\beta^2M(u,\sigma)^\dagger M(u,\sigma))=
\int DF DF^\dagger e^{-\frac{1}{\beta}\sum F^\dagger (M^\dagger M)^{-\frac{1}{2}}F}
\end{equation}   
The key idea of RHMC is to use an optimal (in the minimax sense) rational approximation
to this inverse fractional power.
\[
\frac{1}{x^\frac{1}{2}}\sim \frac{P(x)}{Q(x)}
\] 
where
\[
P(x)=\sum_{i=0}^{N-1}p_ix^i, \quad Q(x)=\sum_{i=0}^{N-1}q_ix^i
\]
Notice that we restrict ourselves to equal order polynomials in numerator and denominator.
In practice it is important to use a partial fraction representation of this rational 
approximation,
\begin{equation}
\frac{1}{x^\frac{1}{2}}\sim \alpha_0+\sum_{i=1}^N\frac{\alpha_i}{x+\beta_i}
\end{equation}
The coefficient $\alpha_i$, $\beta_i$ for $i=1,...,N$ can be computed offline using the Remez
algorithm. Furthermore, the coefficients can be shown to be real positive. Thus the linear systems
are well behaved and, unlike the case of polynomial approximation, the rational fraction 
approximations are robust, stable and converge rapidly with $N$. The resulting pseudofermion
action becomes  
\begin{equation}
S_{PF}=\frac{1}{\beta}\left[\alpha_0F^\dagger F+\sum_i^N F^\dagger 
\frac{\alpha_i}{M^\dagger M  + \beta_i}F\right]
\end{equation}
In principle this pseudofermion action can now be used in a conventional HMC algorithm to yield
an \emph{exact} simulation of the original effective action \cite{Duane}. This algorithm requires that 
we compute the pseudofermion forces. For example, the additional force on the scalar field $u$ due 
to the pseudofermions takes the form,
\[
f_u=\frac{\partial S_{PF}}{\partial u}=-\sum_i^N\alpha_i\chi^{\dagger i}\frac{\partial}
{\partial u}\left( M^\dagger M\right)\chi^i
\] 
where the vector $\chi^i$ is the solution of the linear problem
\[
(M^\dagger M+\beta_i)\chi^i=F
\]
The final trick needed to render this approach feasible is to utilize a \emph{multi-mass} 
solver to solve all $N$ sparse linear systems simultaneously and with a computational cost 
determined primarily by the smallest $\beta_i$. We use a multi-mass version of the usual
Conjugate Gradient (CG) algorithm \cite{Jegerlehner}. Note that for our simulations we have 
used $N=12$ and an approximation that gives us an absolute bound on the relative error of 
$8\times10^{-6}$ for eigenvalues of $M^\dagger M$ ranging from $10^{-6}$ to $10.0$ which 
conservatively covers the range needed. 

\subsection{Spectrum}
\begin{figure}
\centering
\includegraphics[width=7.5cm, angle=-90]{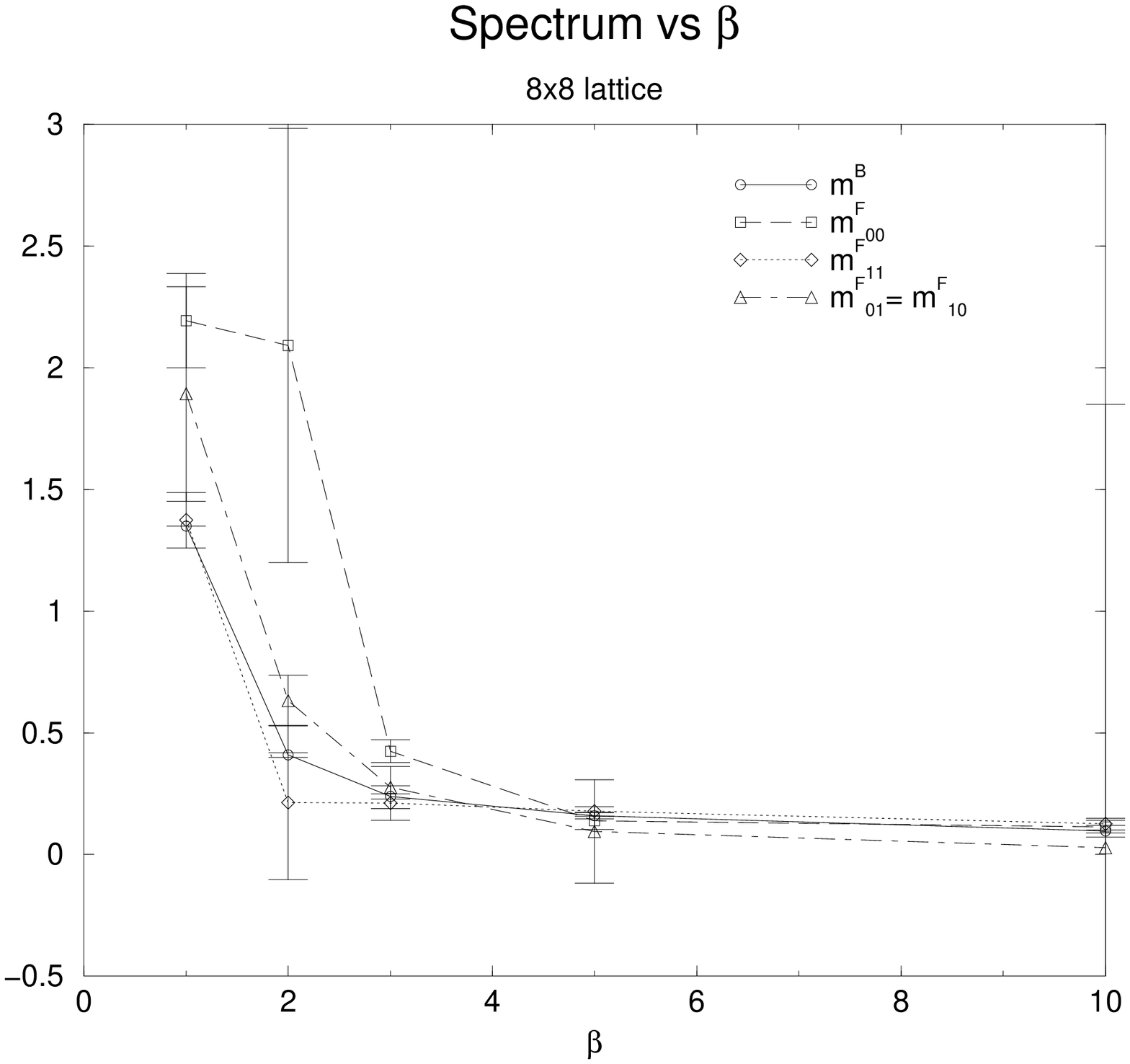}\\
\includegraphics[width=7cm,angle=-90]{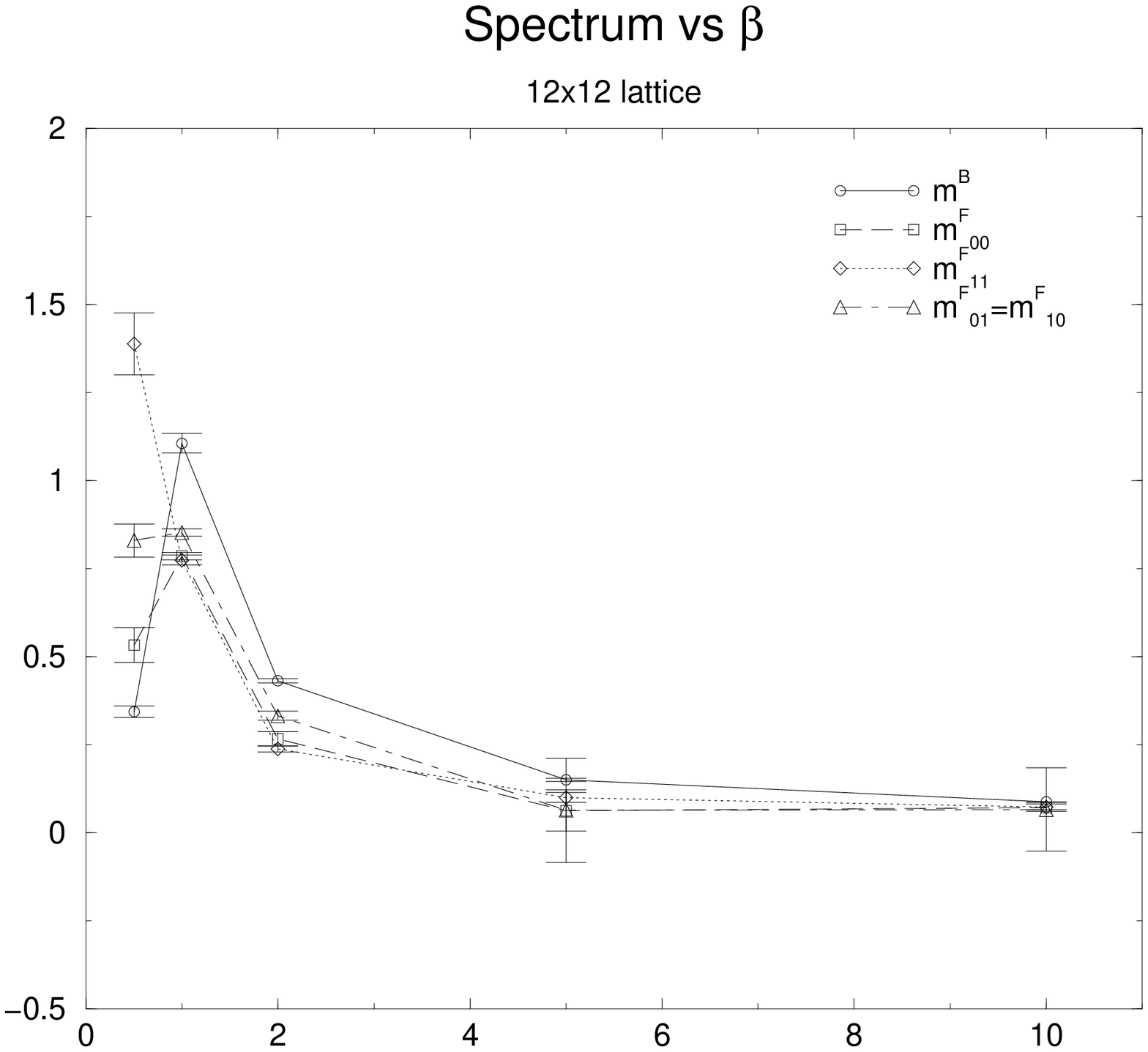}\\
\includegraphics[width=7.5cm,angle=-90]{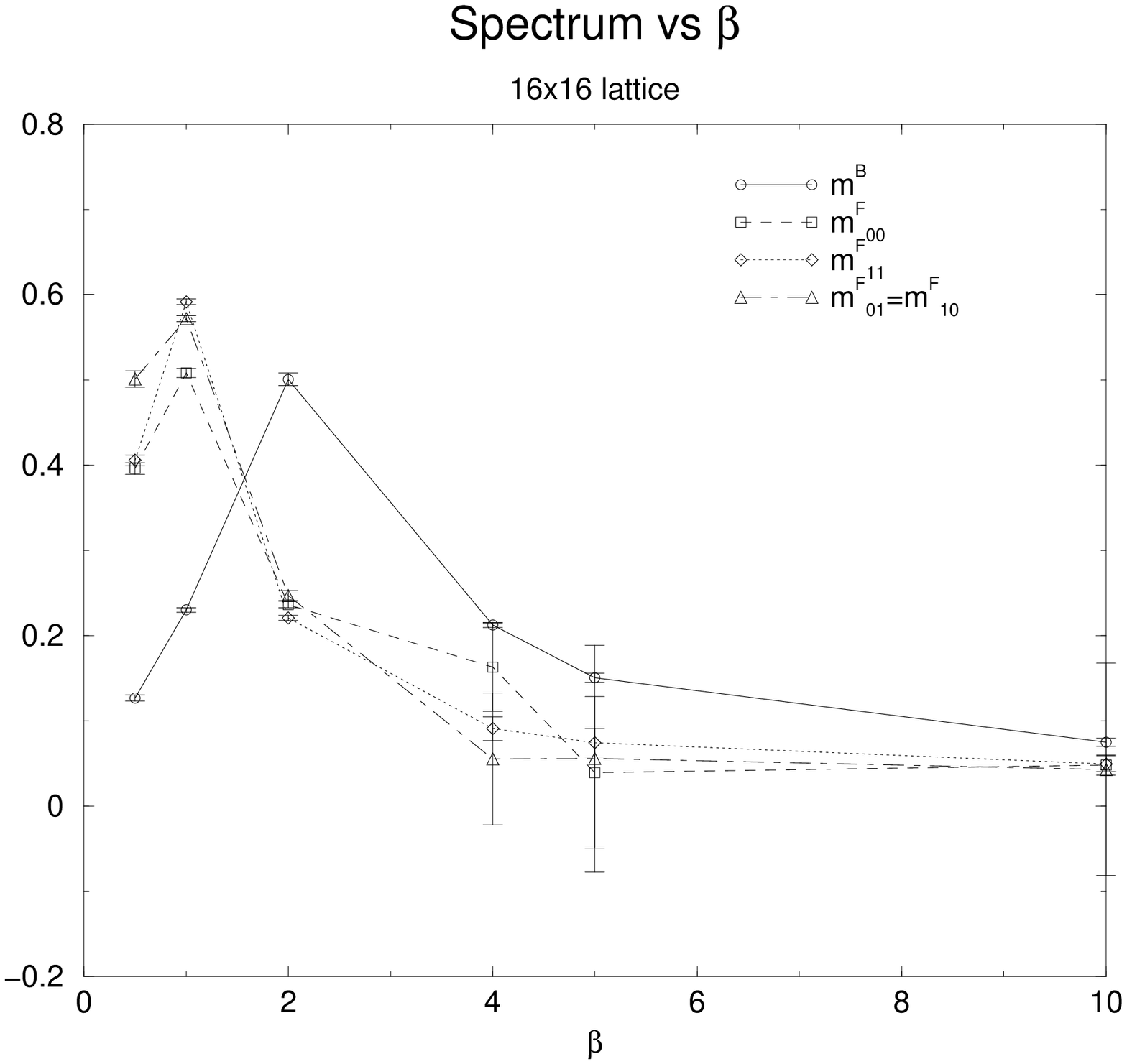}
\caption{mass spectrum for 8x8, 12x12 and 16x16 lattices}
\label{spectrumfig}
\end{figure}

As a check of supersymmetry, we have studied the boson and fermion propagators
projected to zero spatial momentum, namely, $G^B(t)={\rm Re}(<\overline{u}(t)u(0)>)$ and 
$G^F_{ij}(t)=<\overline{\Psi}_i(t)\Psi_j(0)>$. 
One can then extract the lowest lying 
mass states by fitting these two point functions to the form $a+b\cosh(m^B(t-T/2))$ for
bosons and $A(t-T/2)\delta_{ij}+iB(t-T/2)\epsilon_{ij}$ for fermions respectively.
The quantities $A$ and $B$ are even and odd functions of their
argument and we have 
taken the Dirac gamma matrix in the 
time direction to correspond to the
usual Pauli matrix $\sigma_2$. 
To calculate the fermion masses $m^F_{00}$, 
$m^F_{11}$, $m^F_{01}$ and $m^F_{10}$ we have used the simple functions $a\cosh(m^F_{00}(t-T/2))$,
$a\cosh(m^F_{11}(t-T/2))$, $a\sinh(m^F_{01}(t-T/2))$ and $a\sinh(m^F_{10}(t-T/2))$
to fit ${\rm Re}G^F_{00}(t)$, ${\rm Re}G^F_{11}(t)$, ${\rm Im}G^F_{01}(t)$ and ${\rm Im}G^F_{10}(t)$ respectively. 
Figure \ref{spectrumfig} shows the bosonic and fermionic masses versus coupling $\beta$ in 
the range $0.5$ to $10.0$ for different lattice sizes. While the masses are  
quite different at small couplings, the mass degeneracy required by supersymmetry 
is recovered as we approach the continuum limit corresponding to large coupling $\beta$.

\subsection{Ward Identities}
\begin{figure}
\centering
\includegraphics[width=8cm,angle=-90]{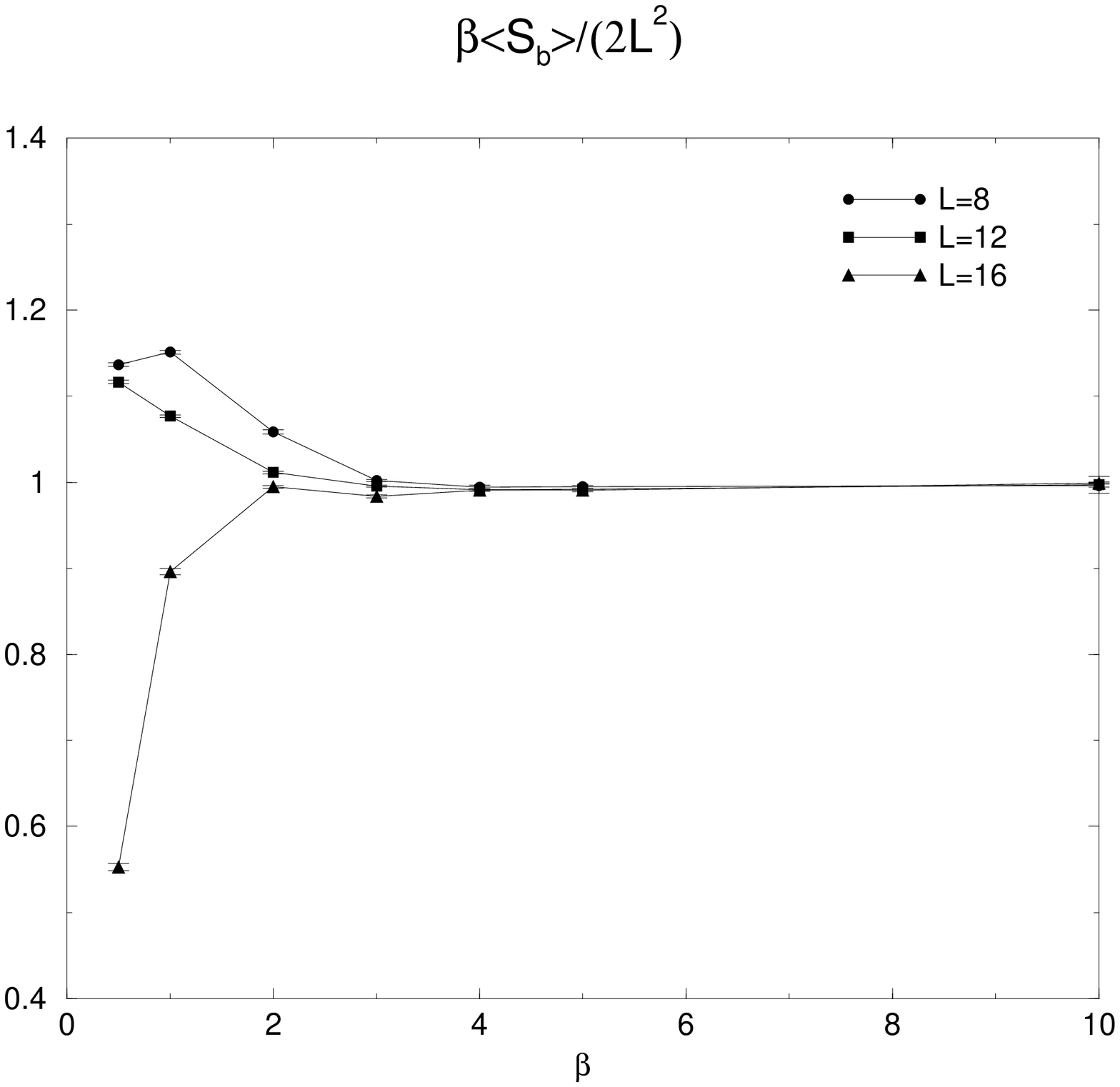}
\caption{Simple Ward Identity}
\label{trivialWI}
\end{figure}

Of prime interest and a stringent test of supersymmetry are the
supersymmetric Ward identities. Consider first the scalar supersymmetry
$Q$.
The Ward identities  
corresponding to $Q$ 
are simply expectation values of the form $<QO>$. The simplest of these corresponds to the action itself, 
indeed as the latter is $Q$-exact \cite{sigma1}, it is clear that,
\begin{eqnarray}
-\frac{\partial\ln Z}{\partial\beta}&=&\frac{1}{Z}\int[D\Phi]e^{-\beta S}S \nonumber\\
&=&\frac{1}{Z}\int[D\Phi]e^{-\beta S}Q\Lambda =<Q\Lambda>\nonumber\\
\end{eqnarray}
which should be zero if the action is
supersymmetric\footnote{Note that, this tells us that 
$Z$ is independent of $\beta$, as long as $\beta$ is not zero, and thus one can evaluate 
$Z$ in the large-$\beta$ limit. Such a limit corresponds to the semi-classical approximation.
Such an approximation is exact in our case (this is true for Witten type theories in 
general)}. On the other hand using 
the effective action in (\ref{effective}),
\begin{equation}
-\frac{\partial\ln Z}{\partial\beta}=0=<S_B>-\frac{2L^2}{\beta}
\end{equation}
Figure \ref{trivialWI} shows a plot of $\frac{\beta}{2L^2}<S_B>$ for a range of couplings 
$\beta$ and different lattice sizes. While there are deviations from
unity at small coupling these seem to 
disappear as the coupling $\beta$ is increased beyond $\beta\sim 4.0$
and yield very strong support to exact $Q$-symmetry in the continuum
limit.  

\begin{figure}
\centering
\includegraphics[width=7.4cm,angle=-90]{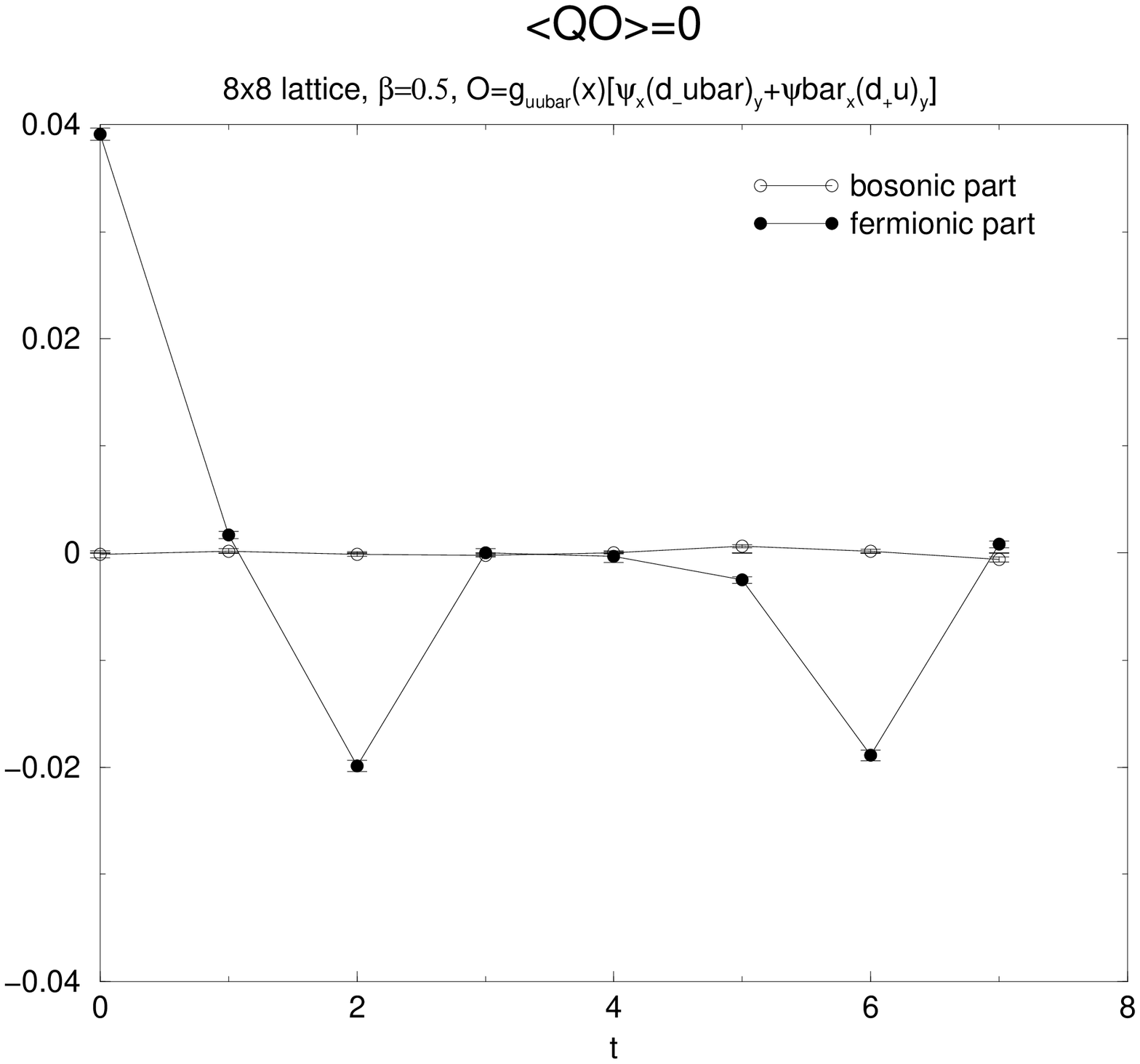}\\
\includegraphics[width=7.4cm,angle=-90]{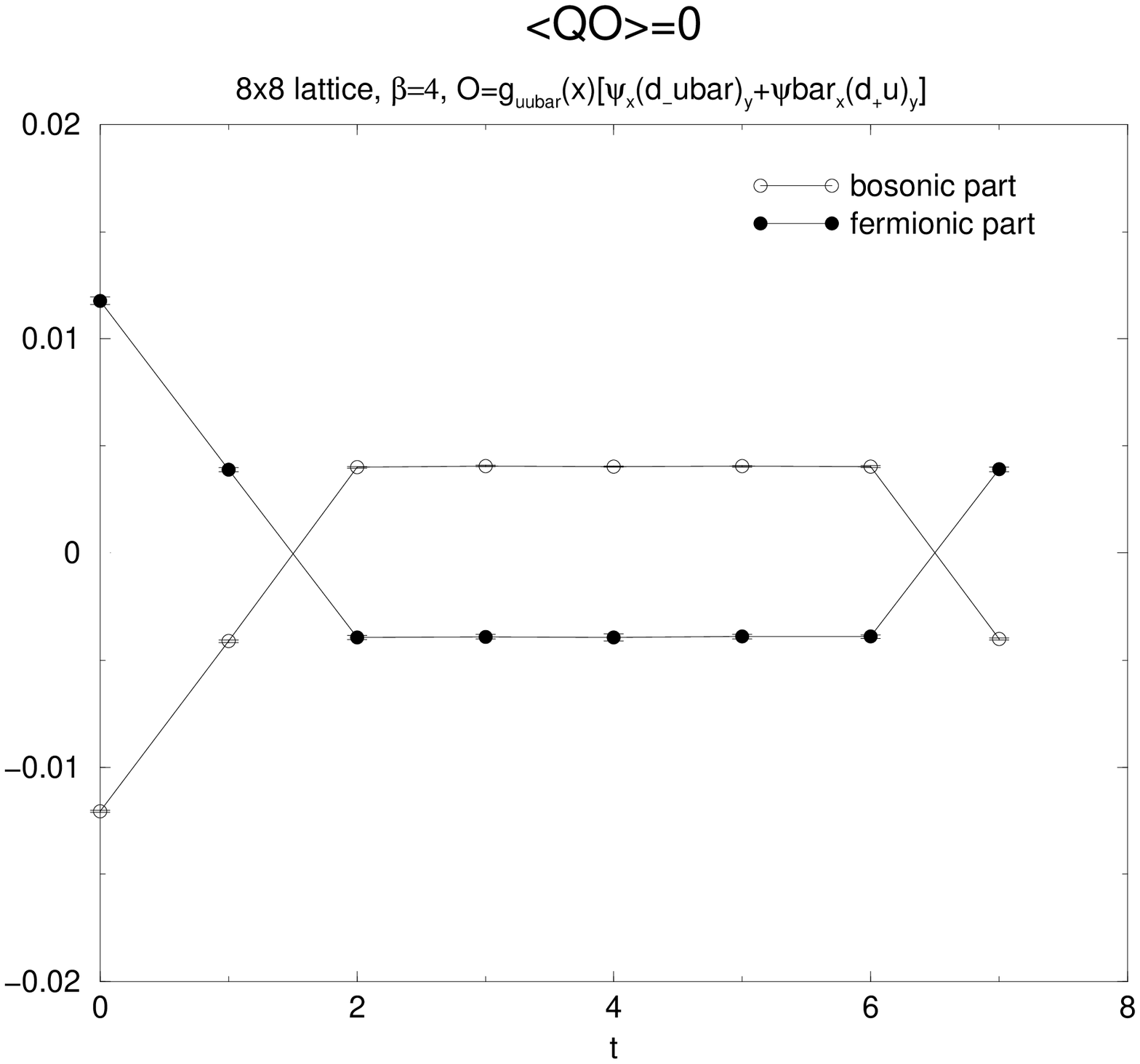}\\
\includegraphics[width=7.4cm,angle=-90]{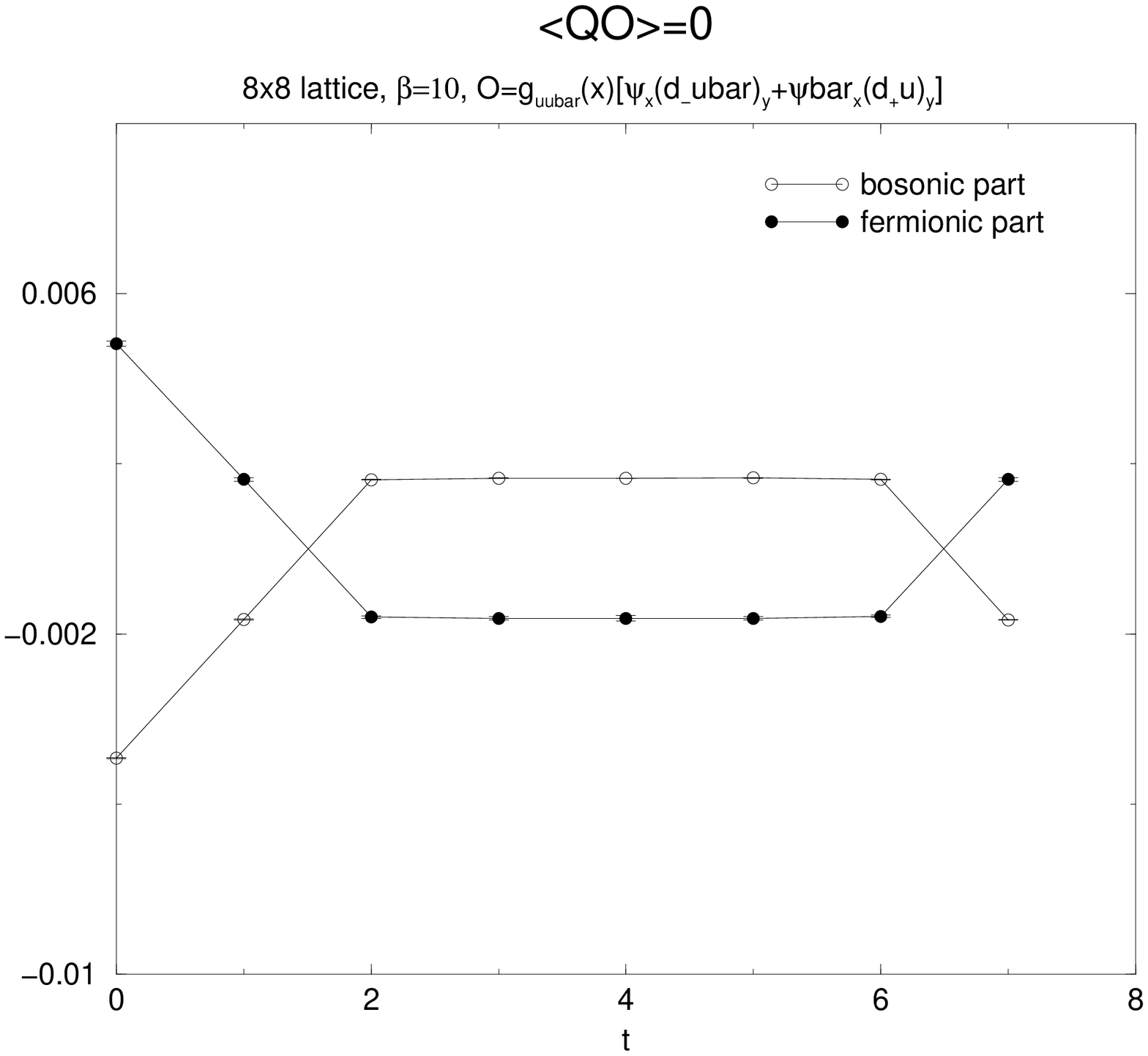}
\caption{$<QO>=0$ for $8\times8$ lattice}
\label{WI1fig8x8}
\end{figure}

\begin{figure}
\centering
\includegraphics[width=7.5cm]{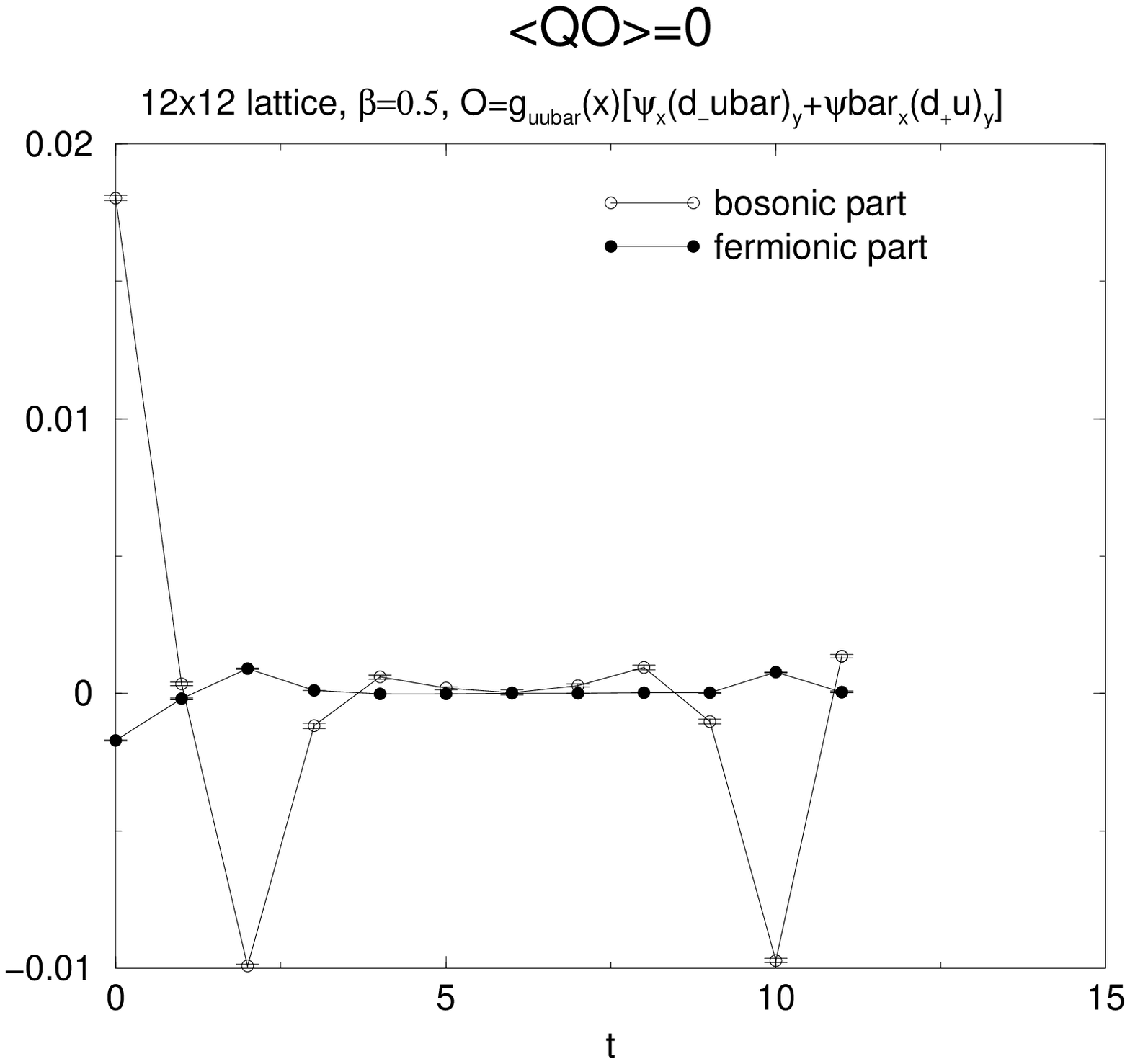}\\
\includegraphics[width=7.5cm]{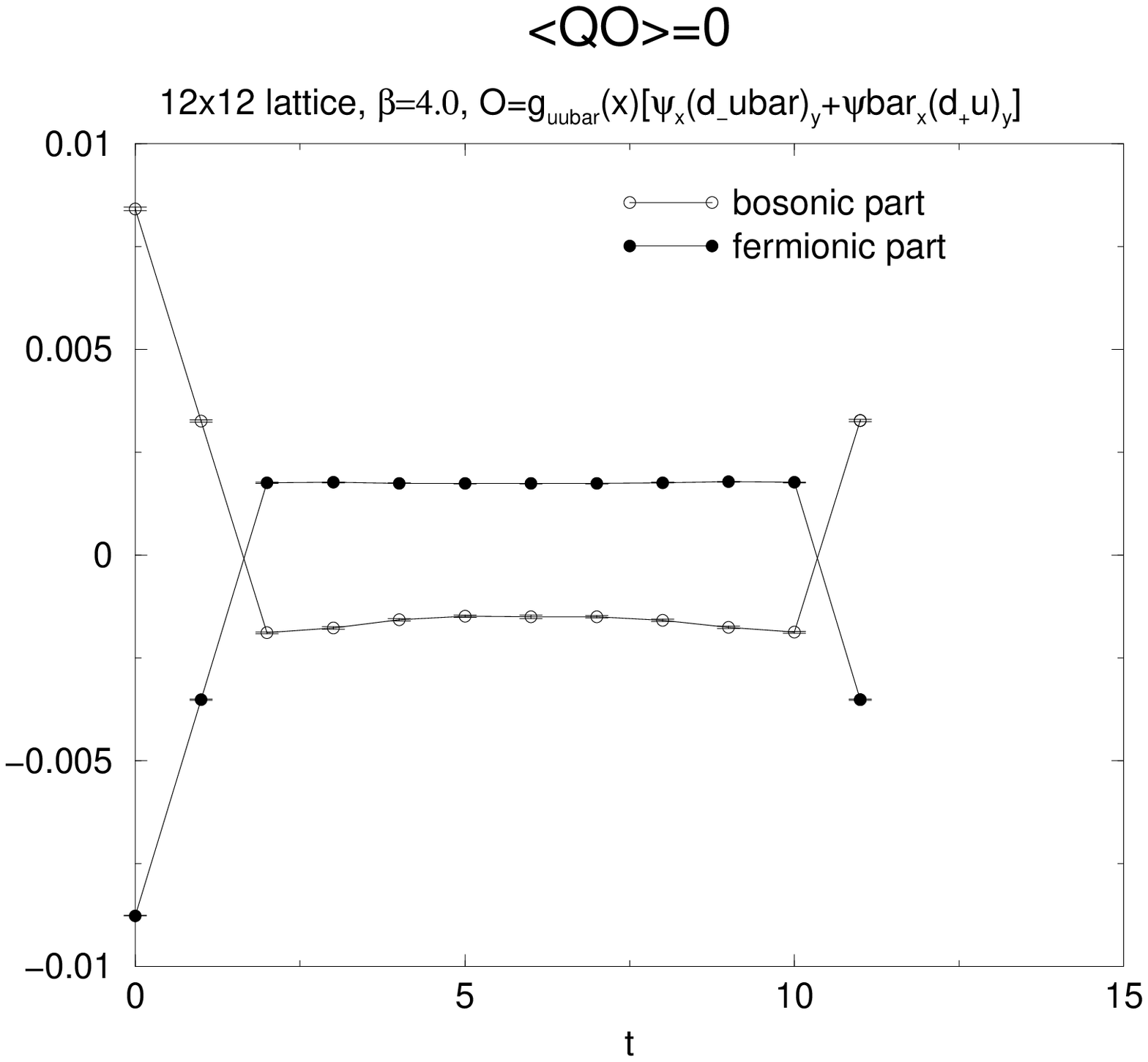}\\
\includegraphics[width=7.5cm]{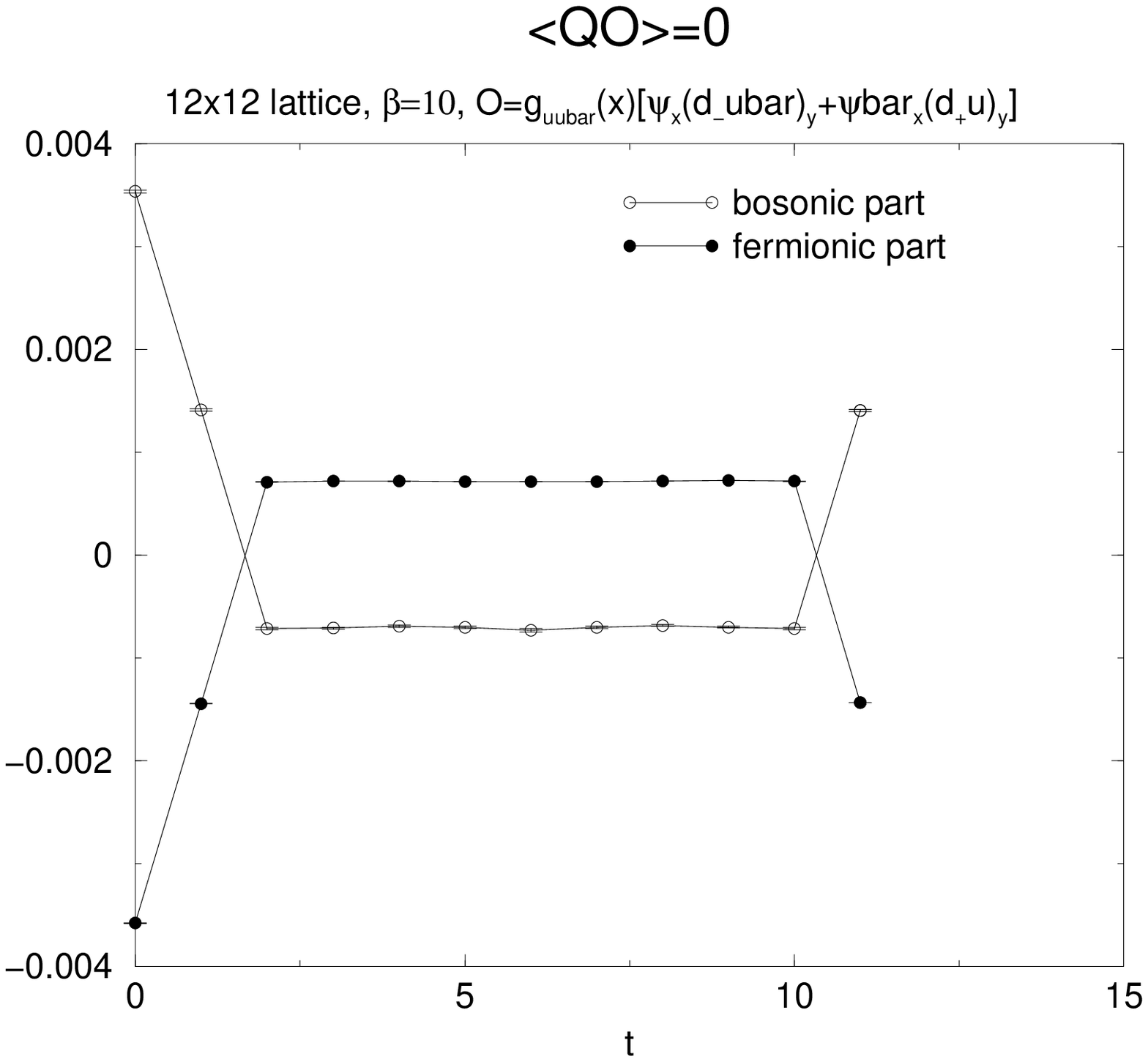}
\caption{$<QO>=0$ for $12\times12$ lattice}
\label{WI1fig12x12}
\end{figure}

\begin{figure}
\centering
\includegraphics[width=7.5cm]{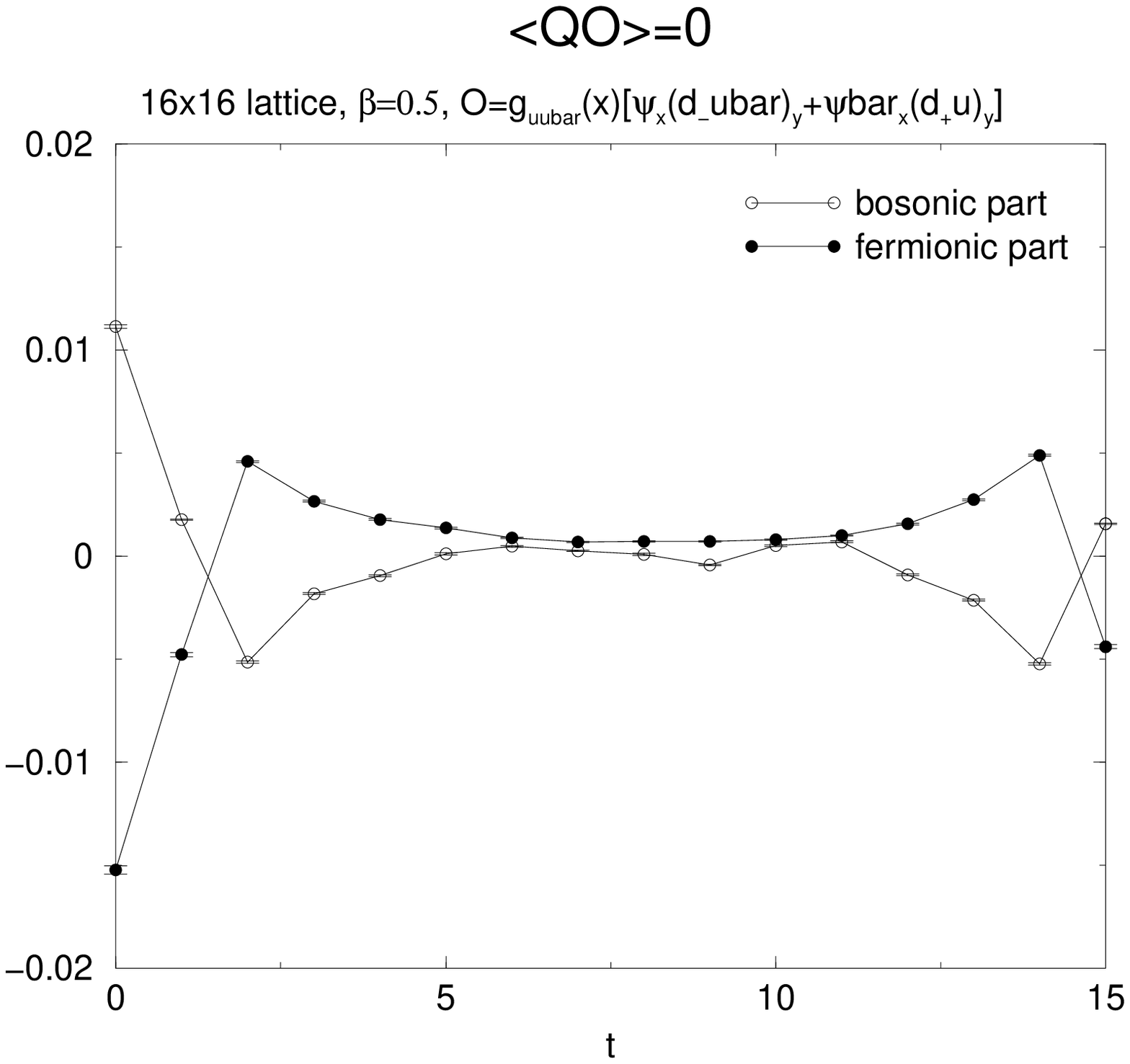}\\
\includegraphics[width=7.5cm]{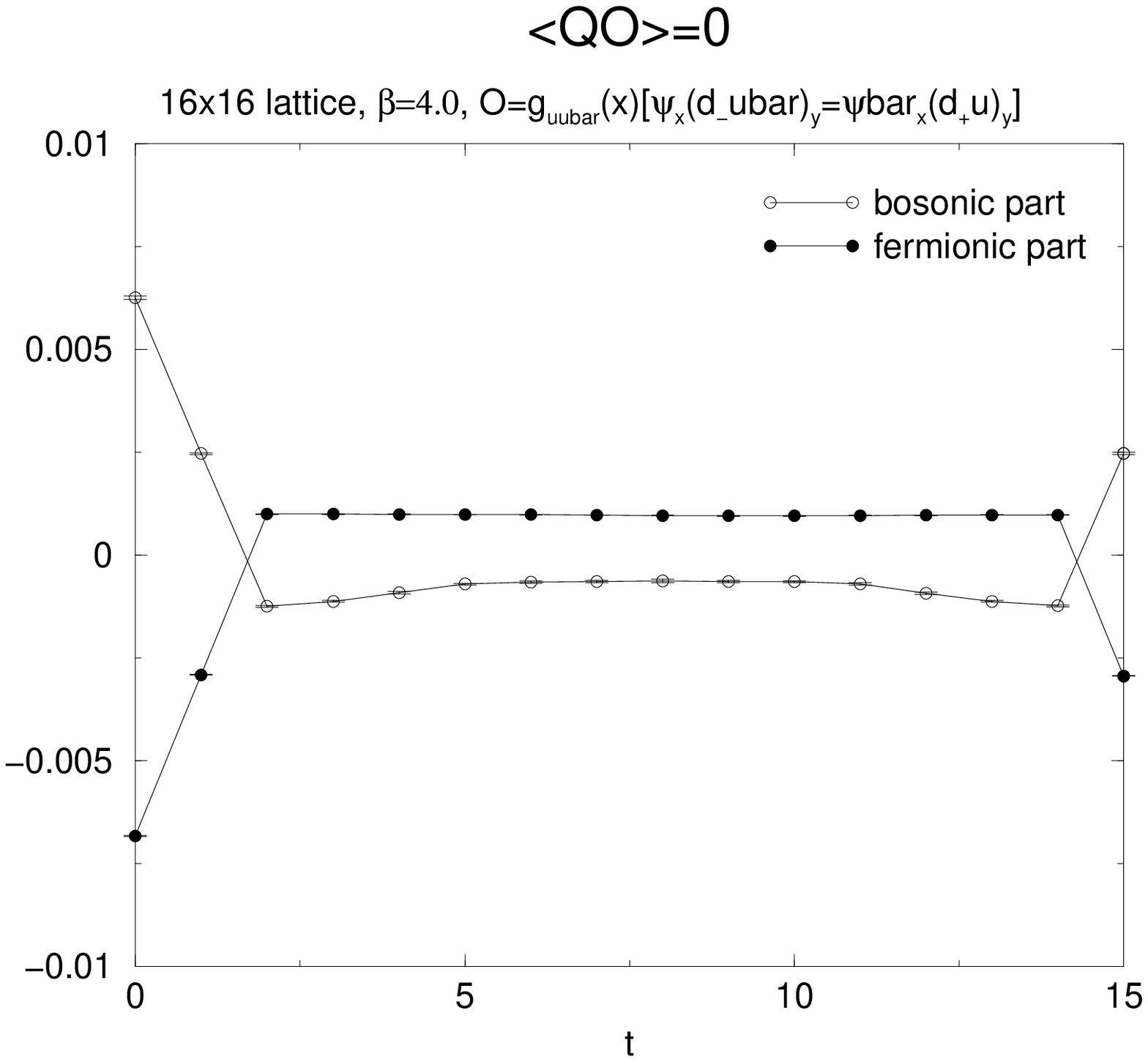}\\
\includegraphics[width=7.5cm]{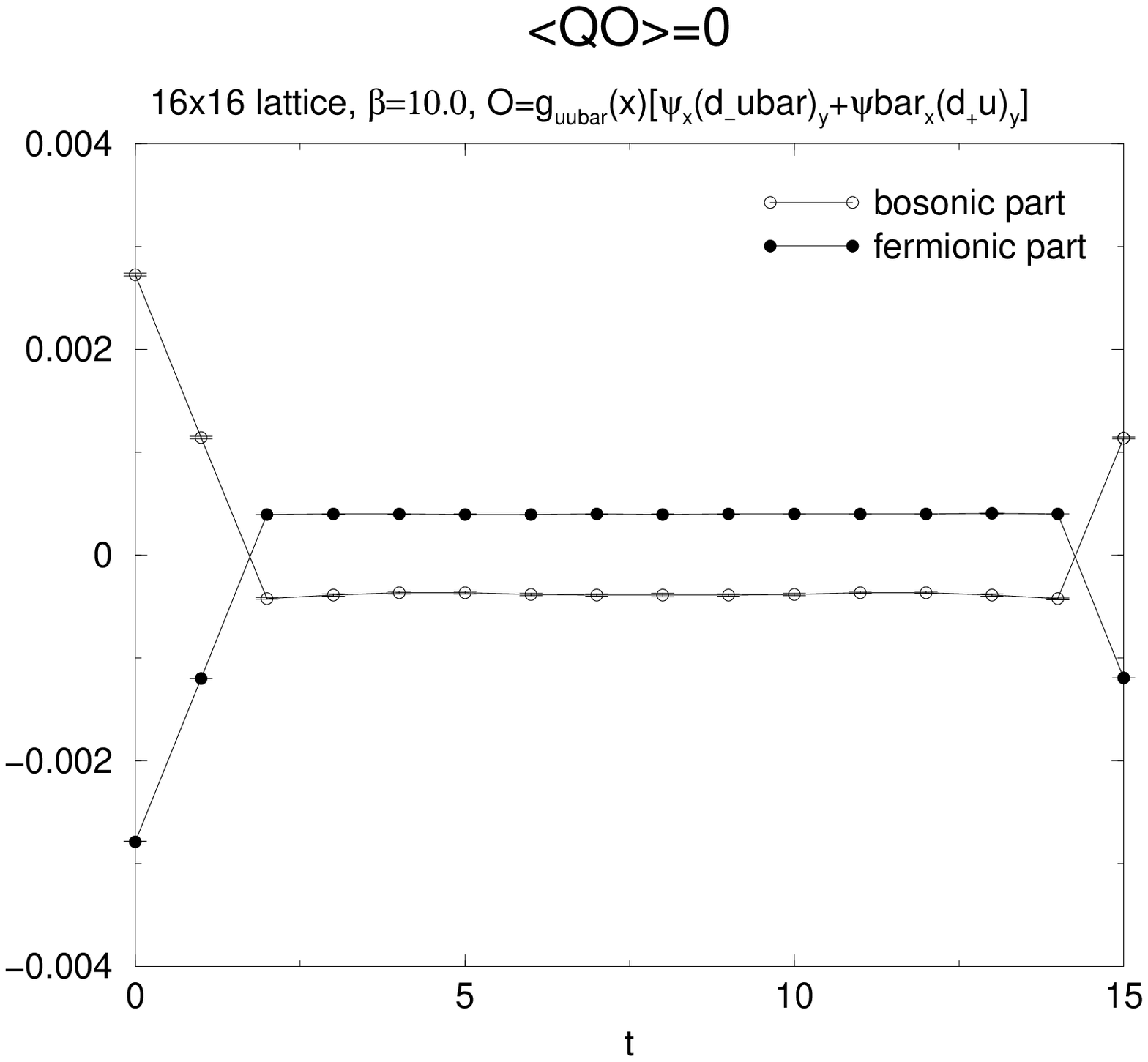}
\caption{$<QO>=0$ for $16\times16$ lattice}
\label{WI1fig16x16}
\end{figure}

We have also looked at other Ward identities. Consider
the local operator $O$ of the form
\begin{equation}
O=h^{+-}g_{I\overline{J}(x)}[\eta^I_{+}(x)\partial_-\phi^{\overline{J}}(y)+
\eta^{\overline{J}}_{-}(x)\partial_+\phi^{I}(y)]
\end{equation}
The operator
$O$ is chosen in such a way that $QO$ is local and leads to a 2-point function
for the fermions. It is also Lorentz invariant in the base space and 
invariant under reparametrizations of the target space (at least in the
continuum limit where the base difference operator becomes a true
derivative). These
constraints ensure the resultant supersymmetric Ward identity is
satisfied non-trivially.
In the $O(3)$ case $<QO>=0$ leads to
\begin{equation}
\label{WI1}
\begin{array}{lcl}
<\frac{1}{2\rho^2(x)}\partial_+u(x)\partial_-\overline{u}(y)>+
<\frac{1}{2\rho^2(x)}\partial_+u(y)\partial_-\overline{u}(x)>&=\frac{1}{2}
<\frac{1}{2\rho^2(x)}\eta(x)\partial_-\overline{\psi}(y)>+& \nonumber \\
&\frac{1}{2}<\frac{1}{2\rho^2(x)}\overline{\eta}(x)\partial_+\psi(y)>&
\end{array}
\end{equation}
Figures \ref{WI1fig8x8}, \ref{WI1fig12x12} and \ref{WI1fig16x16} show plots of 
(\ref{WI1}) projected to zero spatial momentum for different lattice sizes and coupling 
$\beta=0.5,4$ and $10.0$ respectively. While for small coupling $\beta$ the Ward identities
are violated they are clearly satisfied for large coupling and
vanishing lattice spacing. Notice the additional, at first sight rather
startling feature; for $\beta\ge 4.0$ the functions are approximately
independent of the coordinate separation $|x-y|$! Actually this
result follows from the $Q$-exactness of the theory. A continuum theory
which is $Q$-exact has an energy momentum tensor 
$T_{\mu\nu}=\frac{\delta S}{\delta
g_{\mu\nu}}$ which is
also $Q$-exact. Furthermore, it then follows that
the correlation function of two $Q$-invariant operators is
independent of the metric and hence also
of their separation
\cite{Thompson}. Using the $Q$-variation of the
$\eta_+$ field in eqn.~(\ref{Qtrans}) and the equation of motion
$B_+=\partial_+u$ we see that 
\beq
\partial_+ u=Q\eta-\frac{2\overline{u}}{\rho}\psi\eta\eeq
The second term becomes small for large coupling $\beta$ since
$u\sim\beta^{-\frac{1}{2}}$ and hence in this limit the correlator is
dominated by the leading $Q$-exact piece yielding the
constant correlator as we observe. Thus the Ward identity
reveals not only that the scalar supersymmetry is restored for
large $\beta$ but that the energy-momentum tensor and action of
the theory are also $Q$-exact as expected from the
continuum theory.

\begin{figure}
\centering
\includegraphics[width=7.3cm,angle=-90]{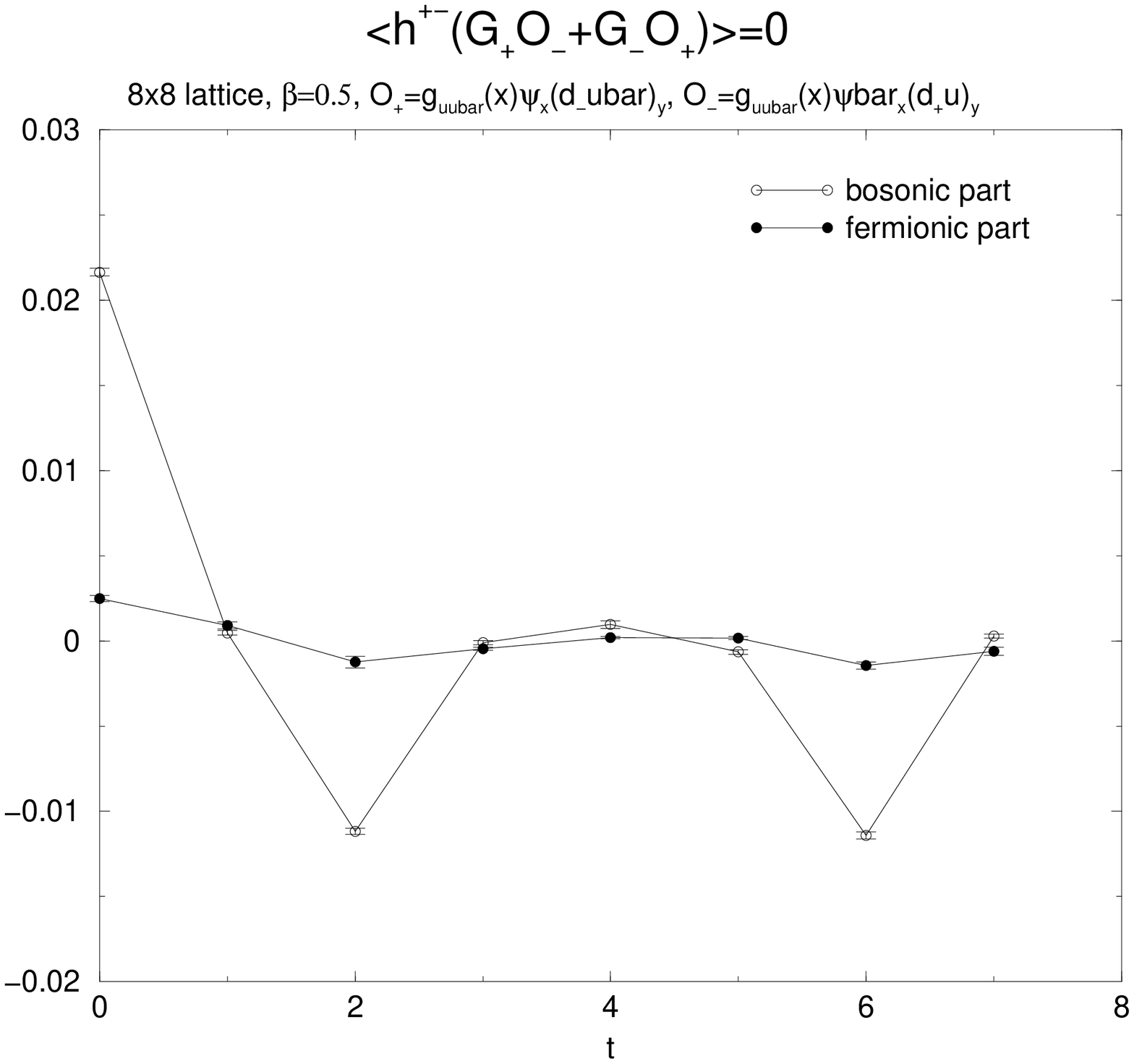}\\
\includegraphics[width=7.3cm,angle=-90]{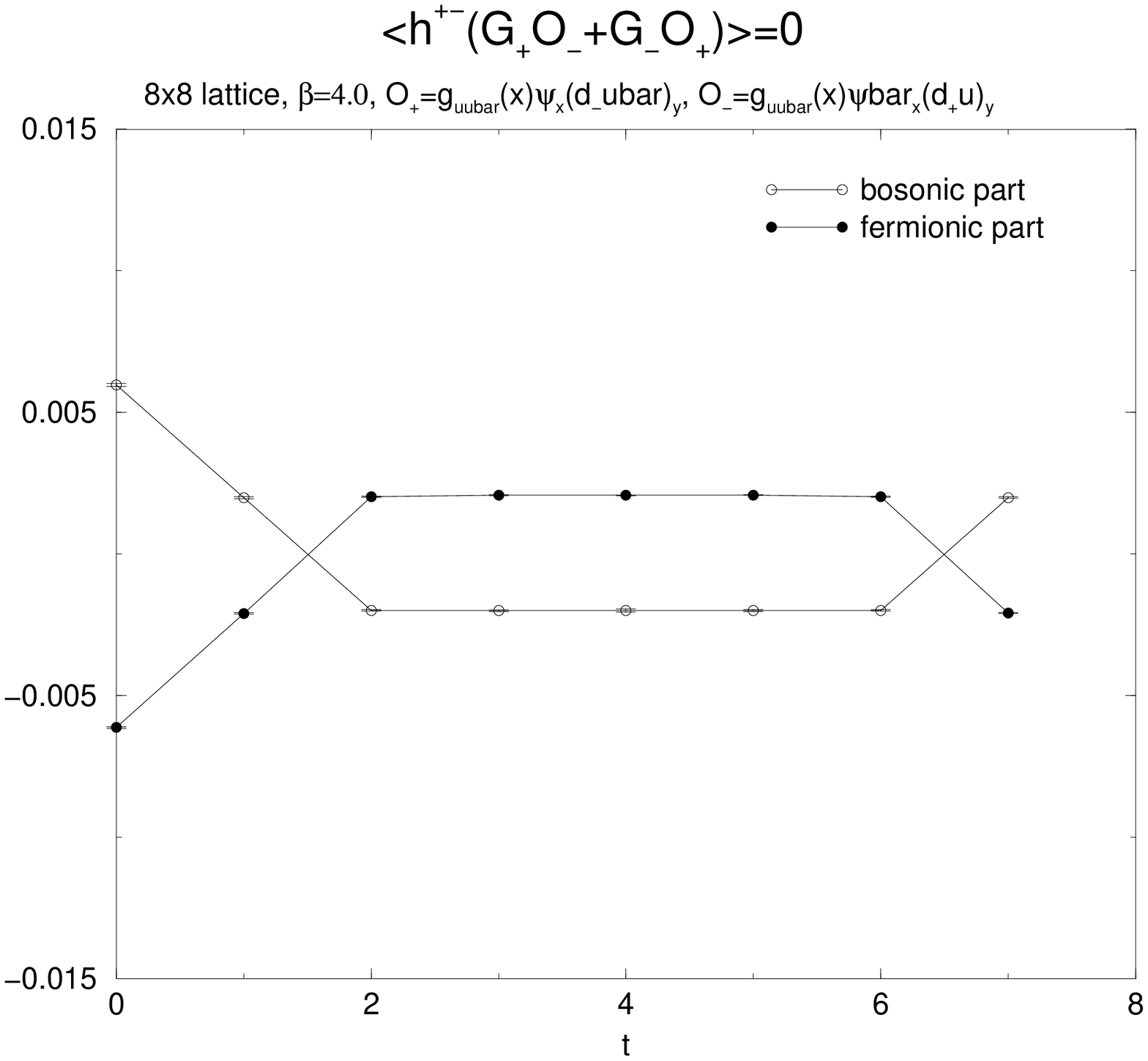}\\
\includegraphics[width=7.3cm,angle=-90]{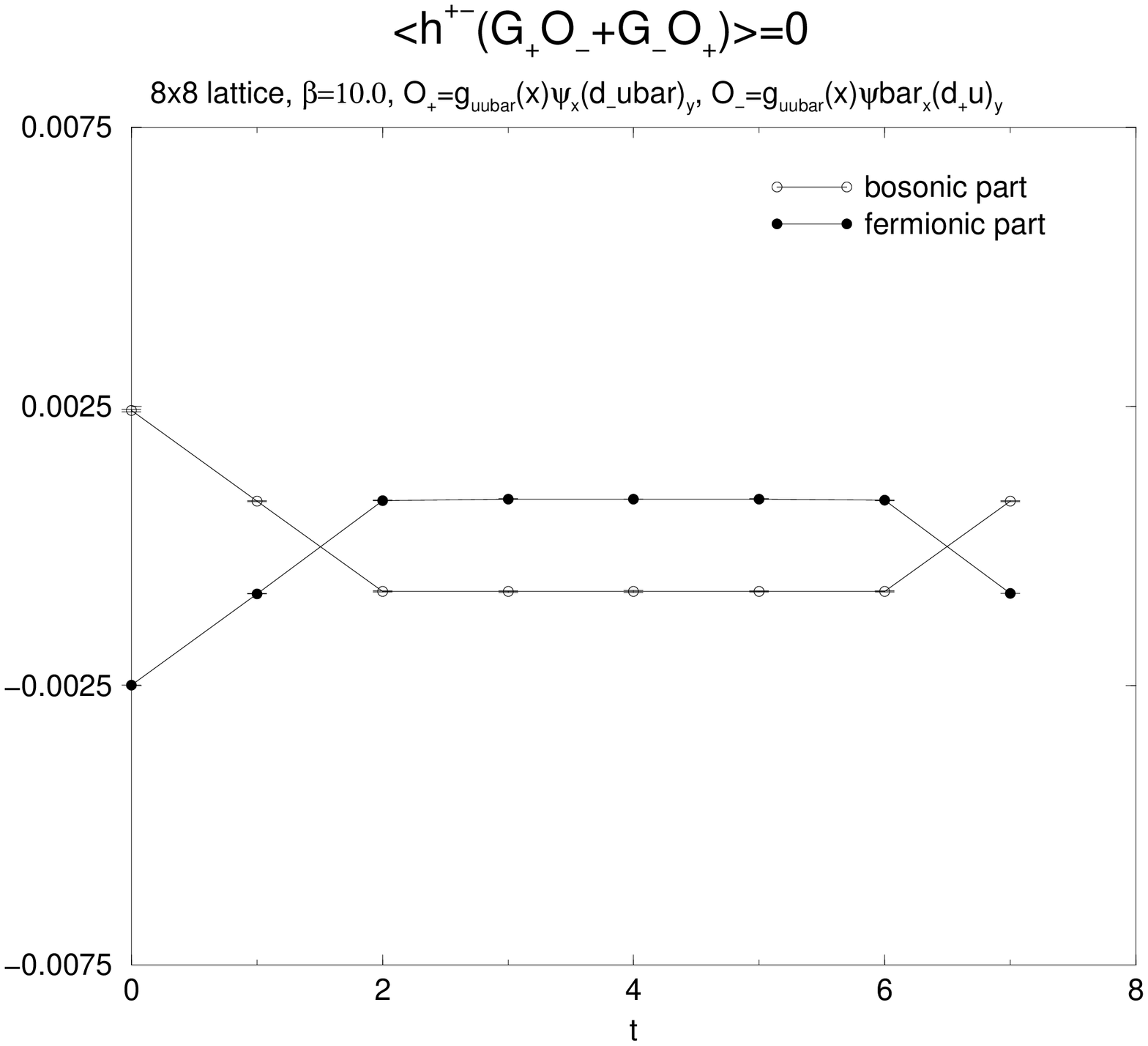}
\caption{$<h^{+-}(G_+O_-+G_-O_+)>=0$ for $8\times8$ lattice}
\label{WI2fig8x8}
\end{figure}

\begin{figure}
\centering
\includegraphics[width=8cm]{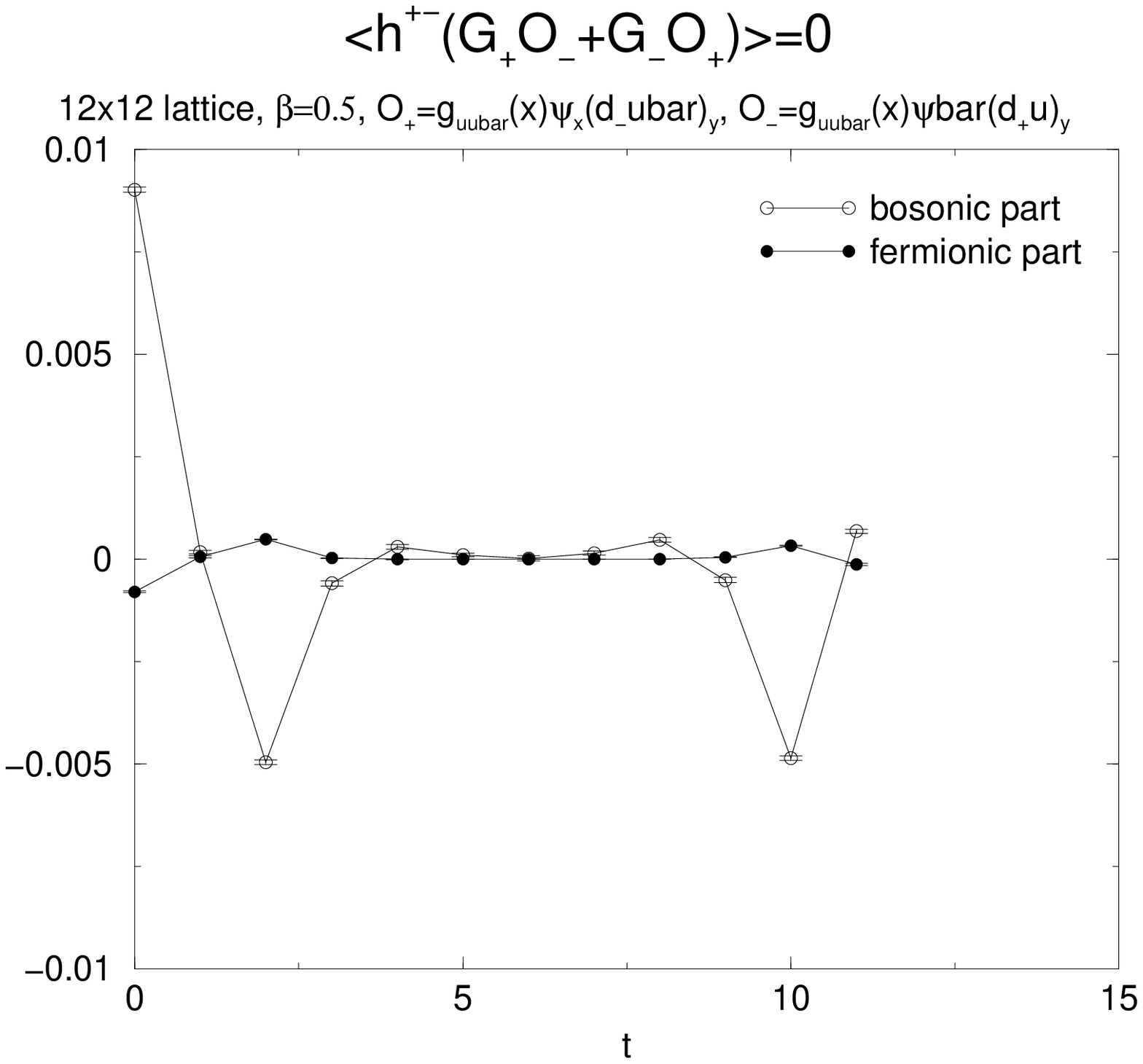}\\
\includegraphics[width=8cm]{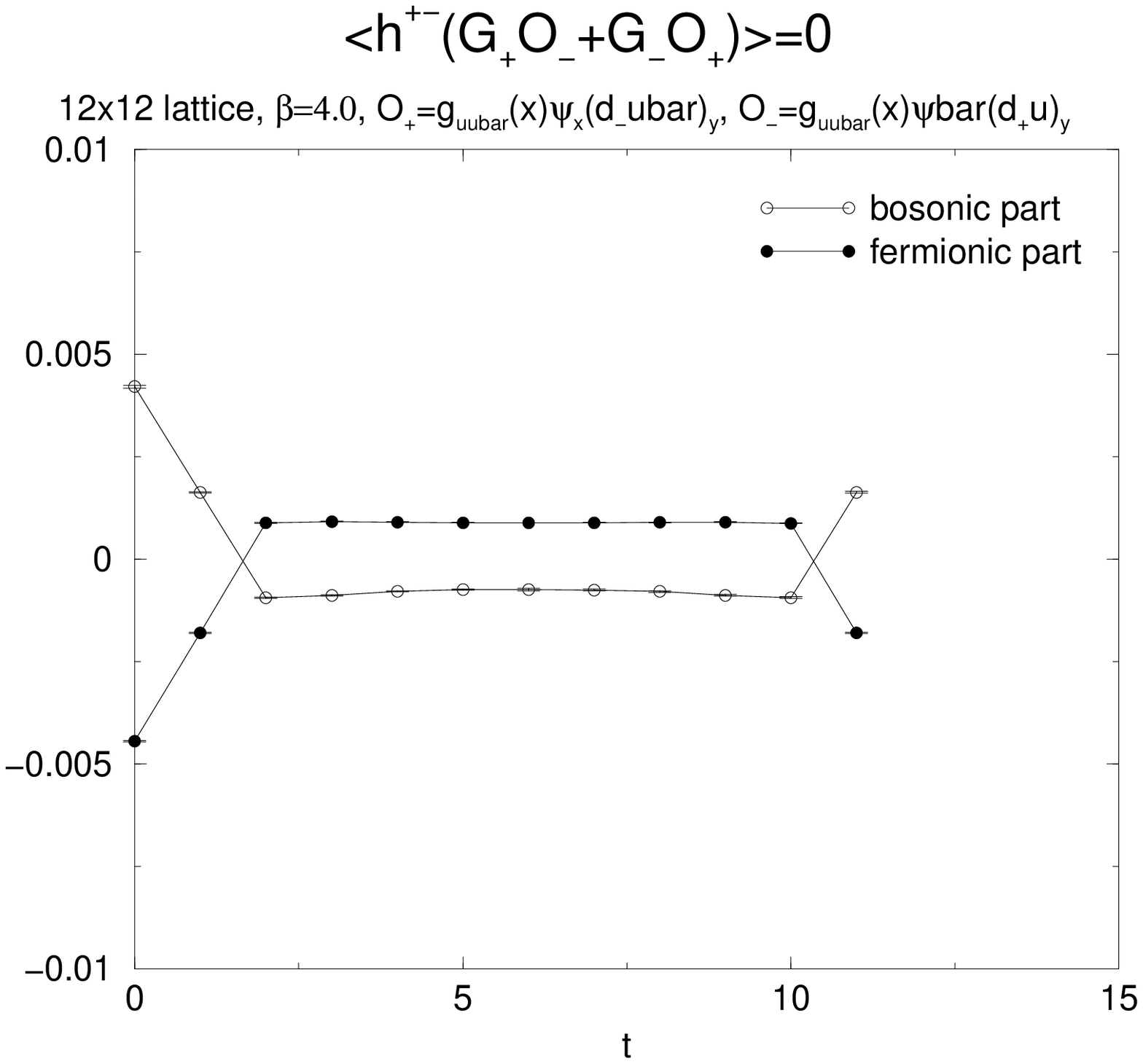}\\
\includegraphics[width=8cm]{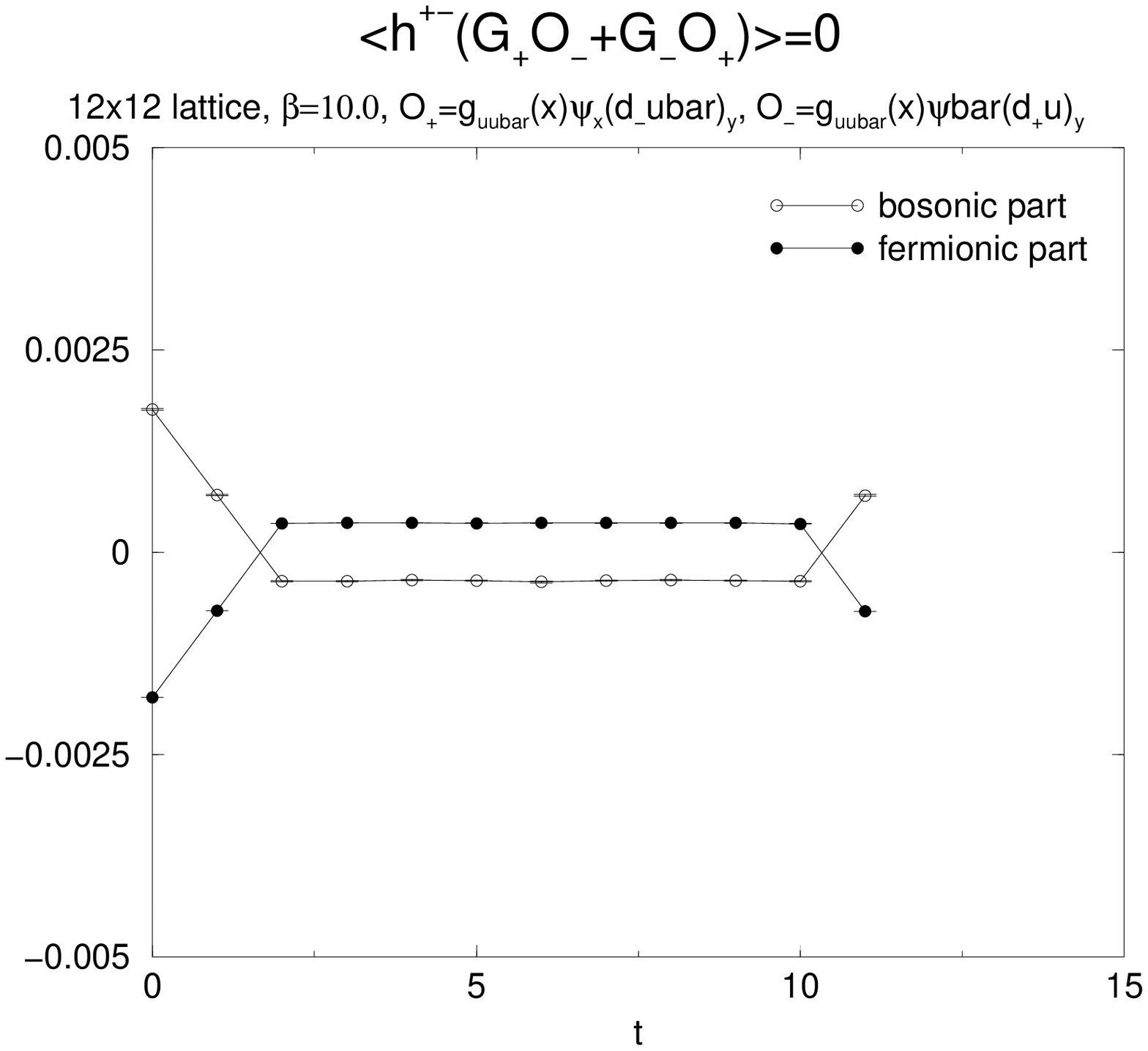}
\caption{$<h^{+-}(G_+O_-+G_-O_+)>=0$ for $12\times12$ lattice}
\label{WI2fig12x12}
\end{figure}

\begin{figure}
\centering
\includegraphics[width=8cm]{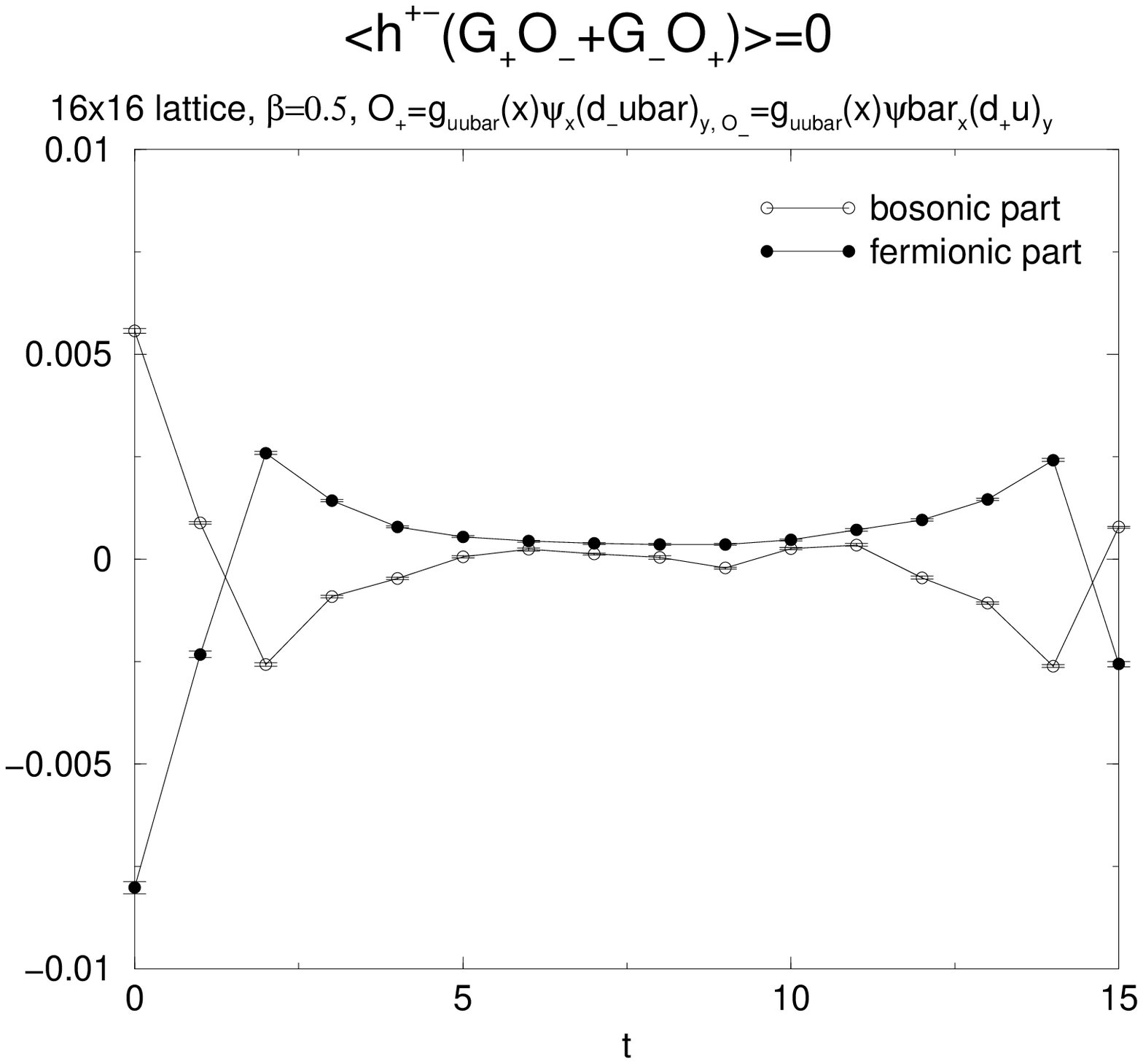}\\
\includegraphics[width=8cm]{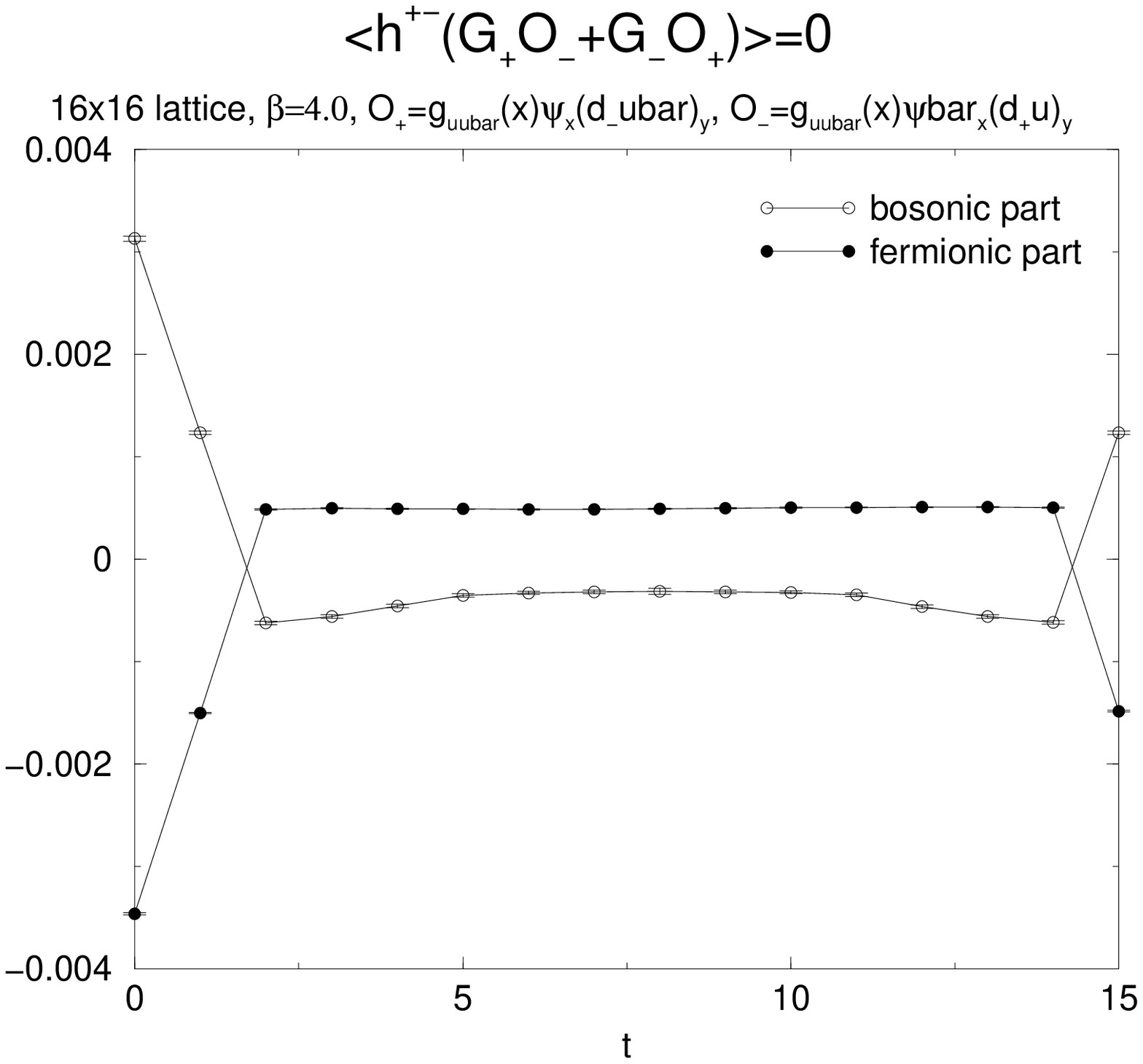}\\
\includegraphics[width=8cm]{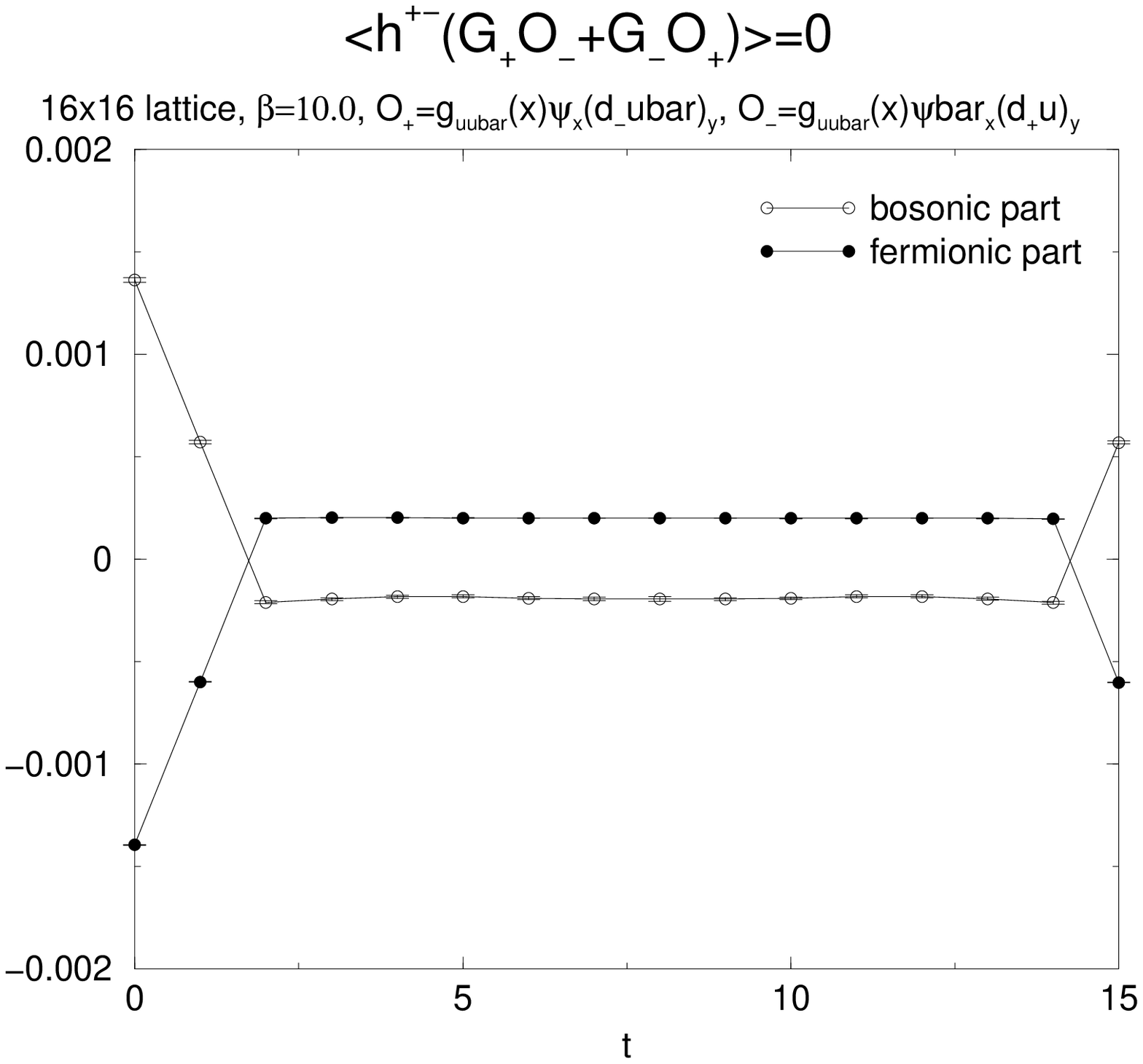}
\caption{$<h^{+-}(G_+O_-+G_-O_+)>=0$ for $16\times16$ lattice}
\label{WI2fig16x16}
\end{figure}

We have also looked at the Ward identities 
following from the other supercharges of the 
twisted $N=2$ supersymmetry (see Appendix \ref{twisted}
for the action of these other supersymmetries). Indeed for the vector 
supercharges $G_\pm$ we have studied 
\begin{equation}
<h^{+-}(G_+O_-+G_-O+)>=0
\label{GO}
\end{equation}
where,
\begin{eqnarray}
O_-=g_{I\overline{J}}(x)\psi^I(x)\partial_-\phi^{\overline{J}}(y)\\
O_+=g_{I\overline{J}}(x)\psi^{\overline{J}}(x)\partial_+\phi^{I}(y)
\end{eqnarray} 
For $O(3)$, eqn.~(\ref{GO}) gives,
\begin{equation}
<\frac{1}{2\rho^2(x)}\partial_+u(x)\partial_-\overline{u}(y)>=
-\frac{1}{2}<\frac{1}{2\rho^2(x)}\overline{\psi}(x)\partial_+\eta(y)>
\label{WI2}
\end{equation} 
Figures \ref{WI2fig8x8}, \ref{WI2fig12x12} and \ref{WI2fig16x16} show plots of 
eqn.~\ref{WI2} projected to zero spatial momentum for different lattice sizes and coupling.
As for the scalar supercharge $Q$ the Ward identities 
due to the vector supercharges are broken for small coupling 
but appear to be restored without fine tuning for small lattice
spacing and large coupling. Again for large enough coupling these
correlation functions appear independent of the temporal coordinate.
This again follows from the fact that asymptotically we
are examining the correlator of two $Q$-exact operators; $Q\eta$ and
$Qu=\psi$.

\begin{figure}
\centering
\includegraphics[width=7.4cm,angle=-90]{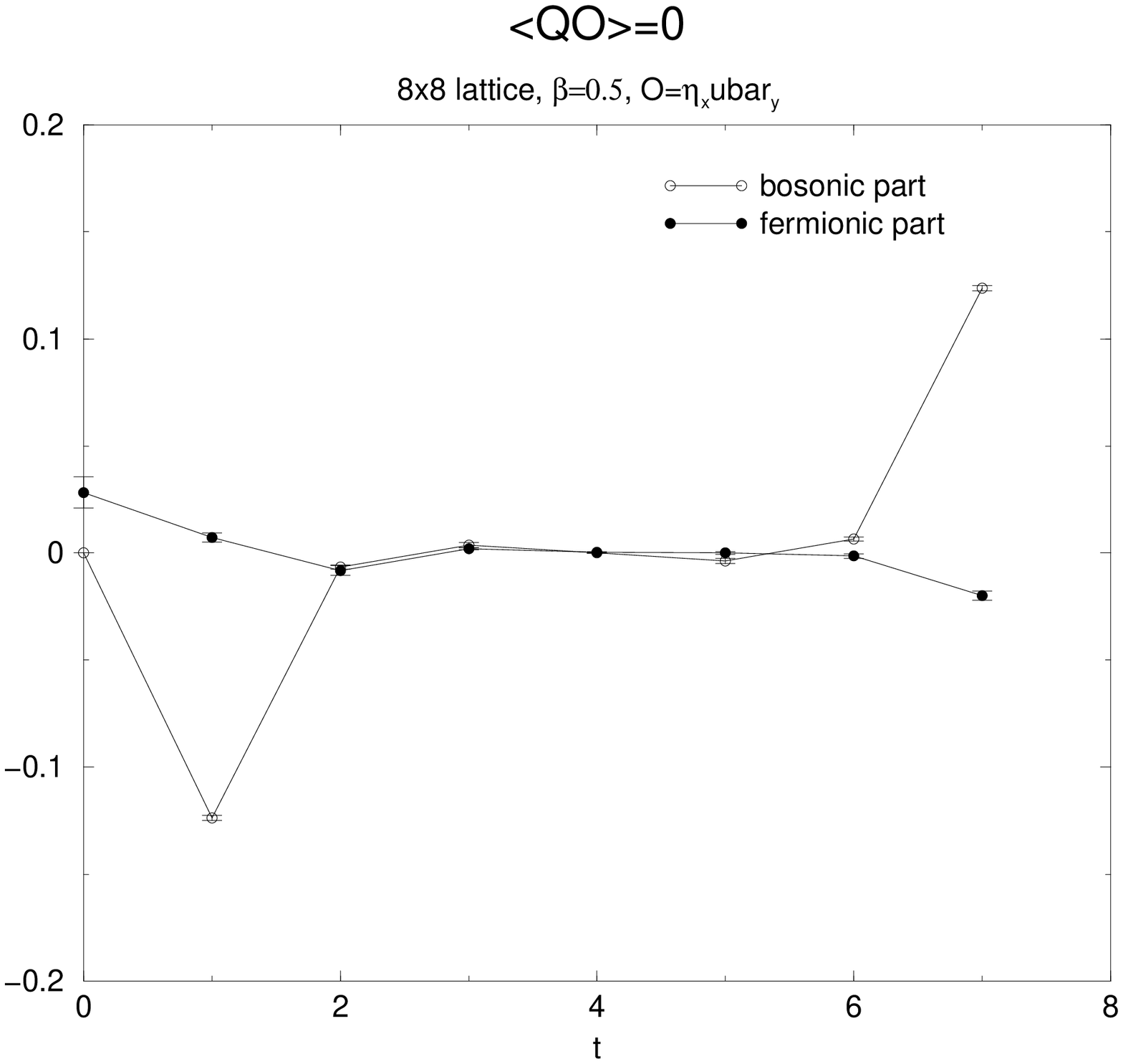}\\
\includegraphics[width=7.4cm,angle=-90]{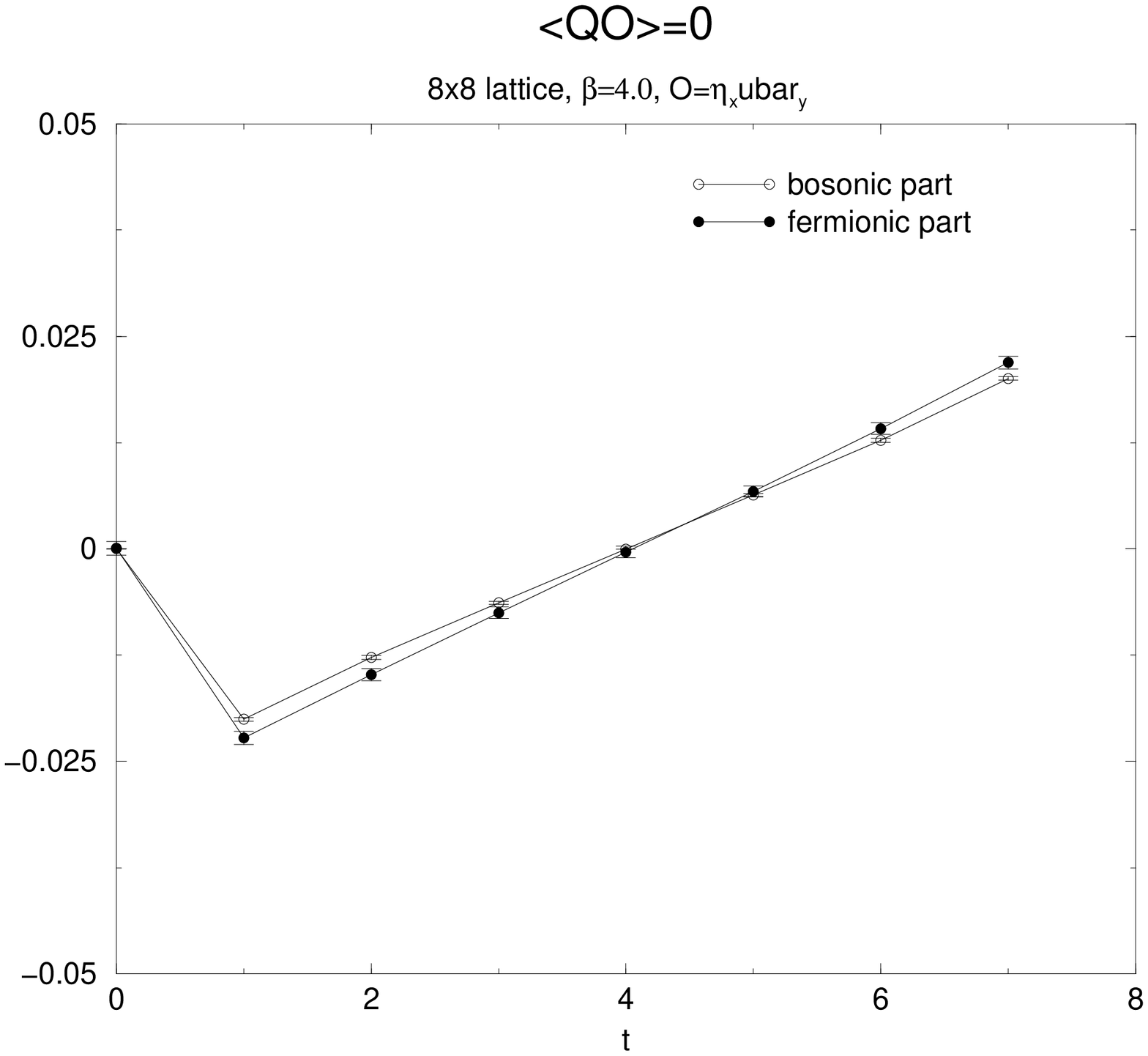}\\
\includegraphics[width=7.4cm,angle=-90]{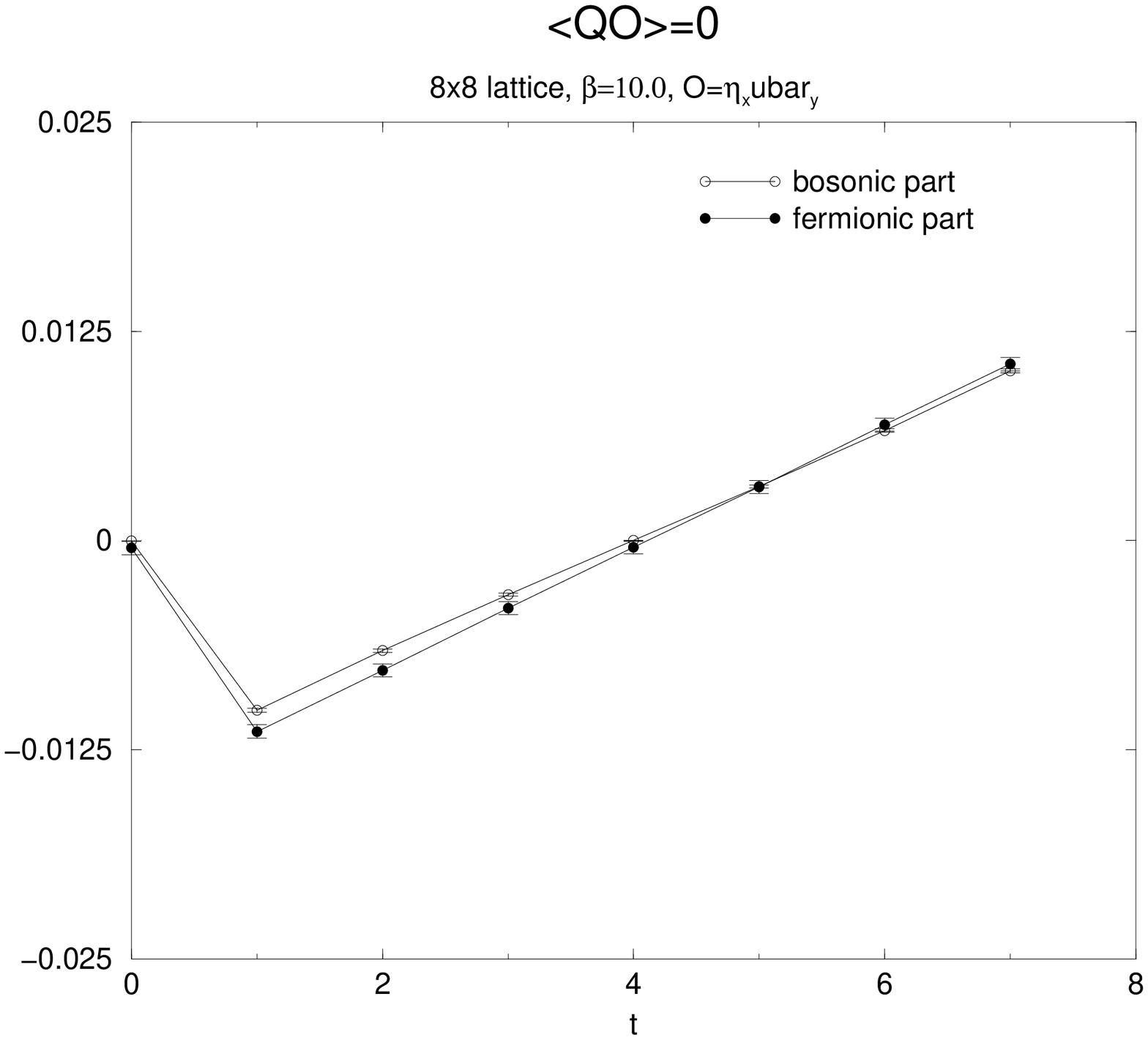}
\caption{$<Q(\eta_x\bar{u}_y)>=0$ for $8\times8$ lattice}
\label{oldWI1fig8x8}
\end{figure}

We have additionally examined a simpler Ward identity. Namely
\begin{equation}
\label{oldQWI}
\frac{1}{2}\langle \eta(x)\bar{\psi}(y)\rangle = \langle \partial_+ u(x) \bar{u}(y)\rangle
\end{equation} 
This Ward identity arises as a result of applying the scalar supercharge $Q$ 
to the operator
$O'=\eta(x)\bar{\phi}(y)$, i.e. $\langle QO'\rangle = 0$.
Figure 
\ref{oldWI1fig8x8} shows a numerical calculation
of eqn. (\ref{oldQWI}) projected to zero spatial momentum. While 
supersymmetry is broken at small $\beta$ coupling it is clearly restored 
as we approach the 
continuum limit, confirming the results we
obtained with the other Ward identities.

\subsection{Fermionic Condensate}
In the previous section we considered Ward identities that lead to 
fermionic 2-point functions. Here we consider Ward identities involving 4-point
functions of the fermion fields. As we will see, the physics here is
richer and more complicated as these quantities can receive
contributions from instantons.
Calculations in the continuum theory
\cite{Novikov,Bohr} predict the presence of a 
vacuum condensate $<\frac{1}{2\rho^2}\overline{\psi}\psi>$   
similar to the gluino condensate in ${\cal N}=1$
super Yang-Mills theory. Since $\psi=\Psi_L$ a single chiral
component of the Dirac field we see that the 
presence of such a condensate
constitutes an anomalous breaking of the chiral symmetry
of the theory.

To see how this comes about in the twisted theory consider the
following correlation function
\beq
C(x-y)=<\Theta(x)\Theta(y)>\eeq
where 
\beq
\Theta(x)=J_{I\overline{J}}\psi^{\overline{J}}\psi^I(x)\eeq
Here, $J_I^J$
is the complex 
structure in the target manifold which locally takes the form $i\delta_I^J$. 
It is straightforward to verify that $\Theta(x)$ is invariant
under $Q$.
Hence, as we argued earlier the correlation function 
$C(x-y)$ should actually be independent of $|x-y|$; $C(x-y)=C$ a
constant. Further, using cluster
decomposition, it can be shown
that any non-vanishing value for $C$ implies a non-vanishing value for
the condensate
$\langle\Theta(x)\rangle=\langle\Theta\rangle=\pm\sqrt{C}$.\footnote{
A word of caution here -- \cite{tim1,tim2} shows that cluster decomposition
{\it fails} in the strong coupling instanton calculation of
the analogous condensate in ${\cal N}=1,D=4$ super Yang-Mills theory.
Here, the topological character of $C$ allows it to be calculated
{\it exactly} at weak coupling}
At first glance the classical
chiral symmetry of the theory would prohibit such a non-zero value
for $C$. However quantum anomalies can and do spoil this symmetry.
As is well known, the theory admits non-trivial
classical solutions called instantons.
These are given by solutions of the equations
\begin{equation}
\partial_+u=0
\end{equation}
The simplest single instanton is hence given by the analytic function
\beq
u(\overline{z})=\frac{\alpha}{z+\beta}\eeq
where $\alpha$ and $\beta$ are complex constants. There are four
real zero modes associated to variation of these parameters and hence,
by supersymmetry, there will be four real fermion zero modes localized
in the vicinity of such an instanton. Examination of the
equation $D_+\Psi=0$ in such a background reveals these
zero modes to be chiral and such a condensation will hence break
chiral symmetry. 
Furthermore, since $\Theta$ is topological
its expectation value can be computed {\it exactly} 
in the semi-classical limit corresponding
to a one loop computation in such an instanton background. 
Notice that the number of fermion zero modes is
just sufficient to saturate the four point function we are
considering. The result
is given in \cite{Novikov}. A lattice calculation of this
quantity is potentially very interesting as it yields information
on both the $U(1)$ anomaly, the presence of approximate lattice
instantons and, as we will see, the dynamical breaking of
supersymmetry. 

\begin{figure}
\centering
\includegraphics[width=7.4cm,angle=-90]{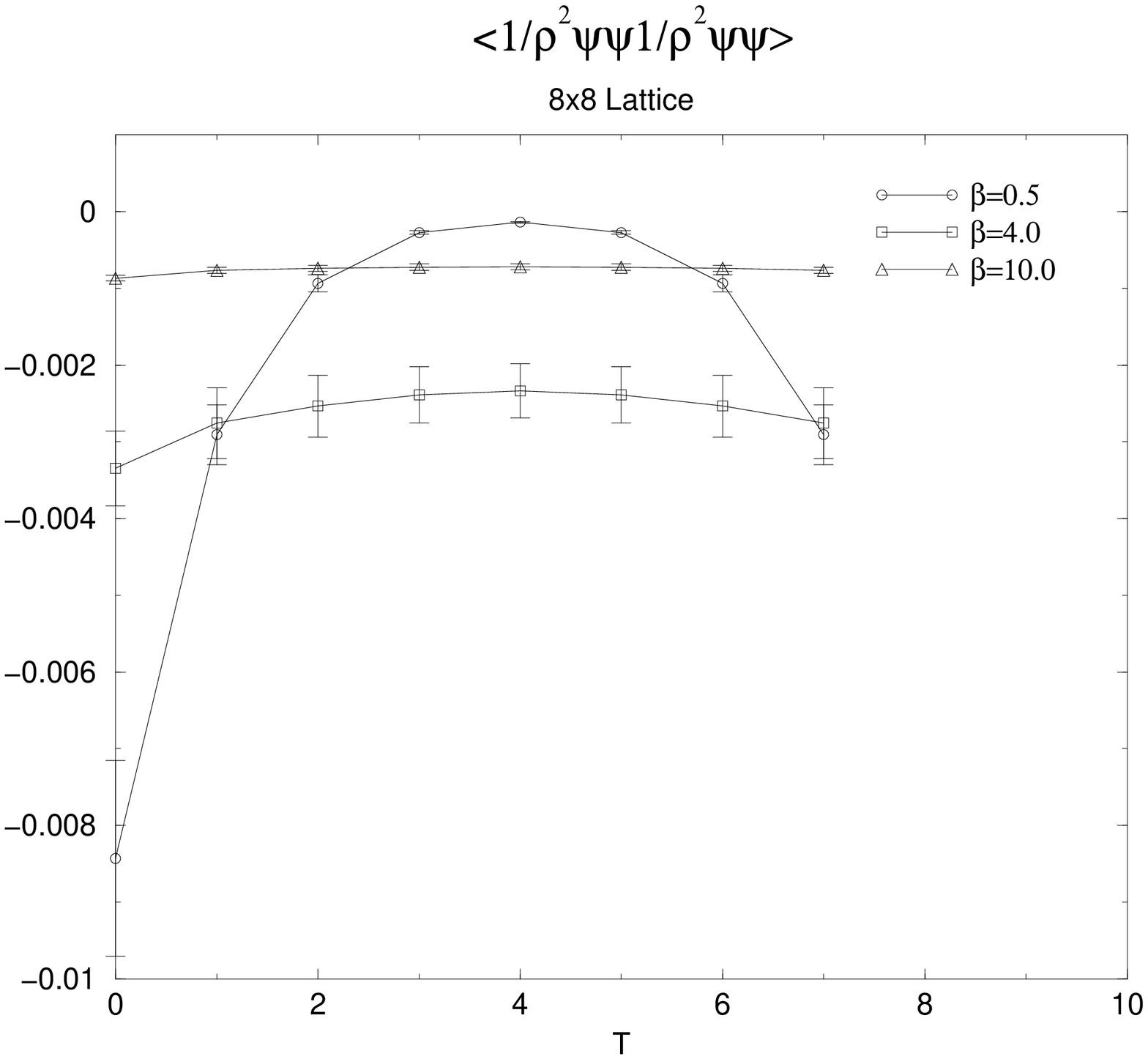}\\
\includegraphics[width=7.4cm,angle=-90]{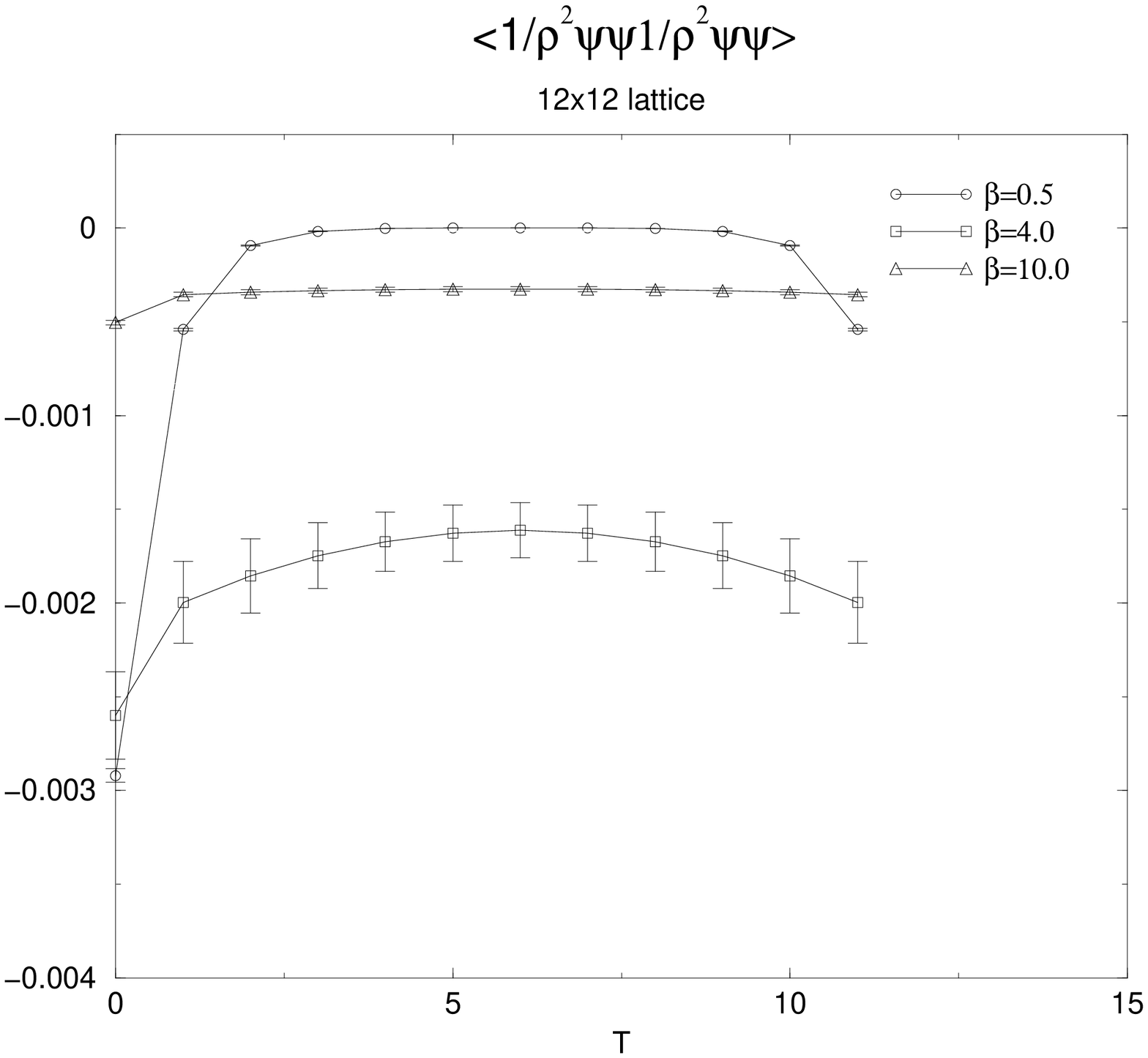}\\
\includegraphics[width=7.4cm,angle=-90]{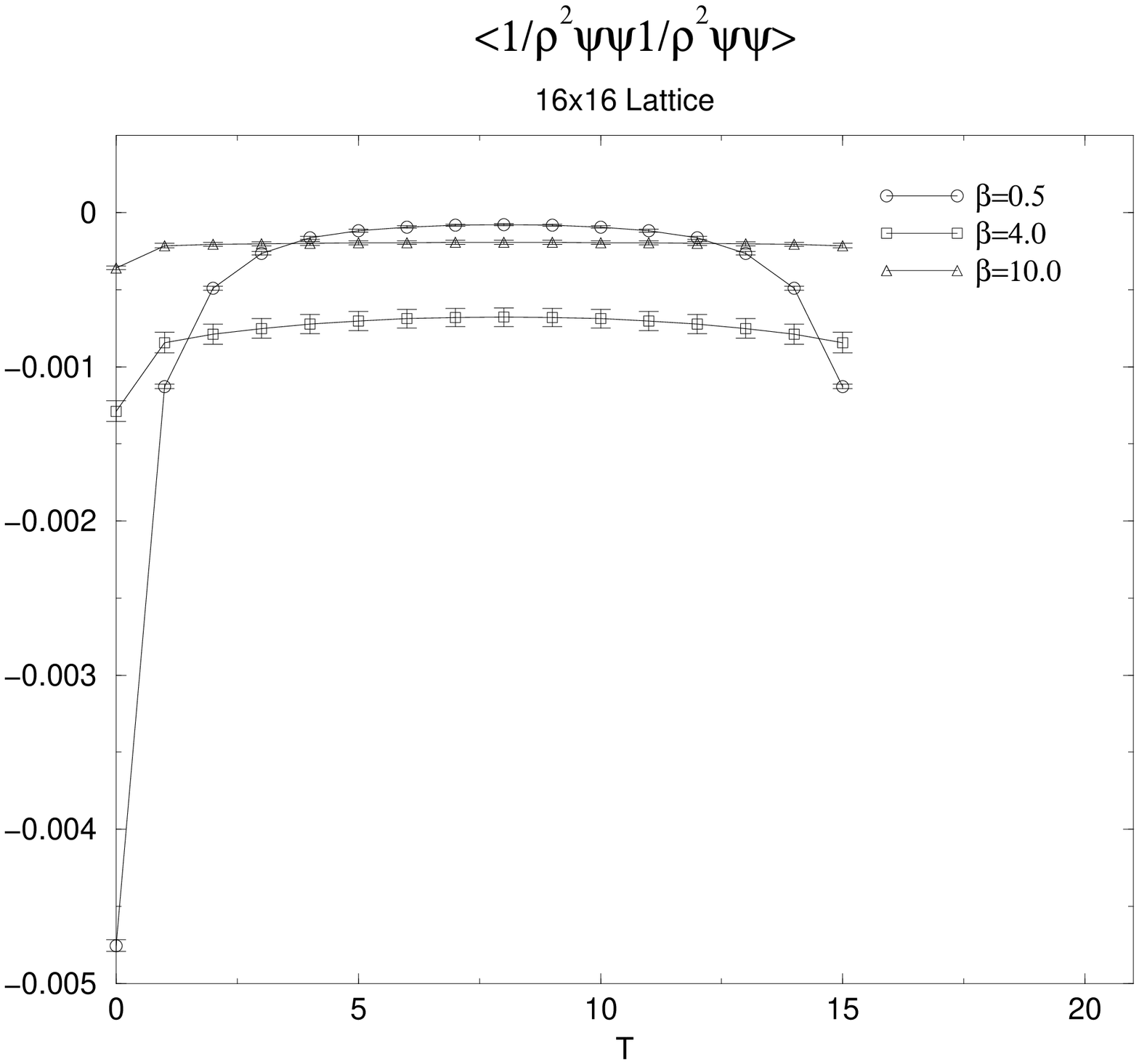}
\caption{condensate for 8x8, 12x12 and 16x16 lattices}
\label{condfig}
\end{figure}

In our lattice regulated $O(3)$ model
we have hence measured the four point function
\begin{equation}
<\Theta>=<\frac{1}{\rho^2(0)}
\overline{\psi}(0)\psi(0)\frac{1}{\rho^2(t)}\overline{\psi}(t)\psi(t)>
\label{cond}
\end{equation}
The plots in figure \ref{condfig}. shows (\ref{cond})
projected to zero spatial momentum for different lattice sizes 
and different values of the 
coupling $\beta$. Clearly in the large $\beta$ limit $<\Theta(t)>$ 
is approximately independent of $t$ confirming
the presence of a condensate. 

Of course on a torus only instanton/anti-instanton pairs can exist but provided
these are well separated and
the discretization errors small 
we can still expect a condensation of four approximate
zero modes in the vicinity of such a lattice instanton yielding a
corresponding contribution to the local
value of $C$. In practice the value of the correlator is not constant
as we will find contributions to $C$ also from the anti-instanton.
This is particularly true for small lattice volumes as can be
seen in the figures \ref{condfig}. There are also explicit
lattice supersymmetry breaking effects visible in the data at $\beta=0.5$.
Notice that the value of the condensate for a fixed
lattice volume initially grows as
$\beta$ is increased but eventually turns over and decreases.
We can understand this as the result of finite volume effects which become
large for large $\beta$ and suppress
such instanton-like configurations.
Notice that the correlation function we plot is dimensionless
and hence contains powers of the lattice spacing $a(\beta)$. We have
not measured this quantity directly so it hard to assess from
figure~\ref{condfig} whether this condensate survives the continuum
limit. We will try to address this question in the next section.

We have argued that the presence of such a condensate is signal
for the anomalous breaking of chiral symmetry but it can also be
viewed as an order parameter for supersymmetry breaking. To see this
notice that the condensate of the
$O(3)$ model can actually be obtained as the
$Q$-variation of the following operator.
\begin{equation}
O_{\rm cond}=\frac{\overline{u}(x)}{\rho(x)\rho^2(y)}\psi(x)\overline{\psi}(y)\psi(y)
\end{equation}
Thus a non-zero value for this quantity is equivalent to
the statement $<QO>\ne 0$
and indicates a dynamical
breaking of supersymmetry driven by instantons rather along the
lines originally envisaged by Witten \cite{witten_break}.
However it was pointed out in
\cite{Novikov,Bohr} that
this operator $O_{\rm cond}$ is {\it not} invariant under the global
$O(3)$ symmetry of the model. It is conjectured that
supersymmetry is only violated in the {\it unphysical}
$O(3)$ non-invariant
sector of the theory. Specifically, the expectation is
that any operator $QO$ containing a multiple of a four-fermion term 
and for 
which $O$ is not $O(3)$ invariant would develop a 
vacuum expectation value different from zero whereas
for an $O(3)$ invariant $O$ it would remain zero. For a general ${\cal N}=2$
sigma model the notion of $O(3)$ invariance would then be replaced by
the requirement that the operator be invariant under all possible
isometries of the target metric. It would be interesting to
investigate these issues in more detail.

\subsection{Phase}
\begin{figure}
\centering
\includegraphics[width=10cm,angle=0]{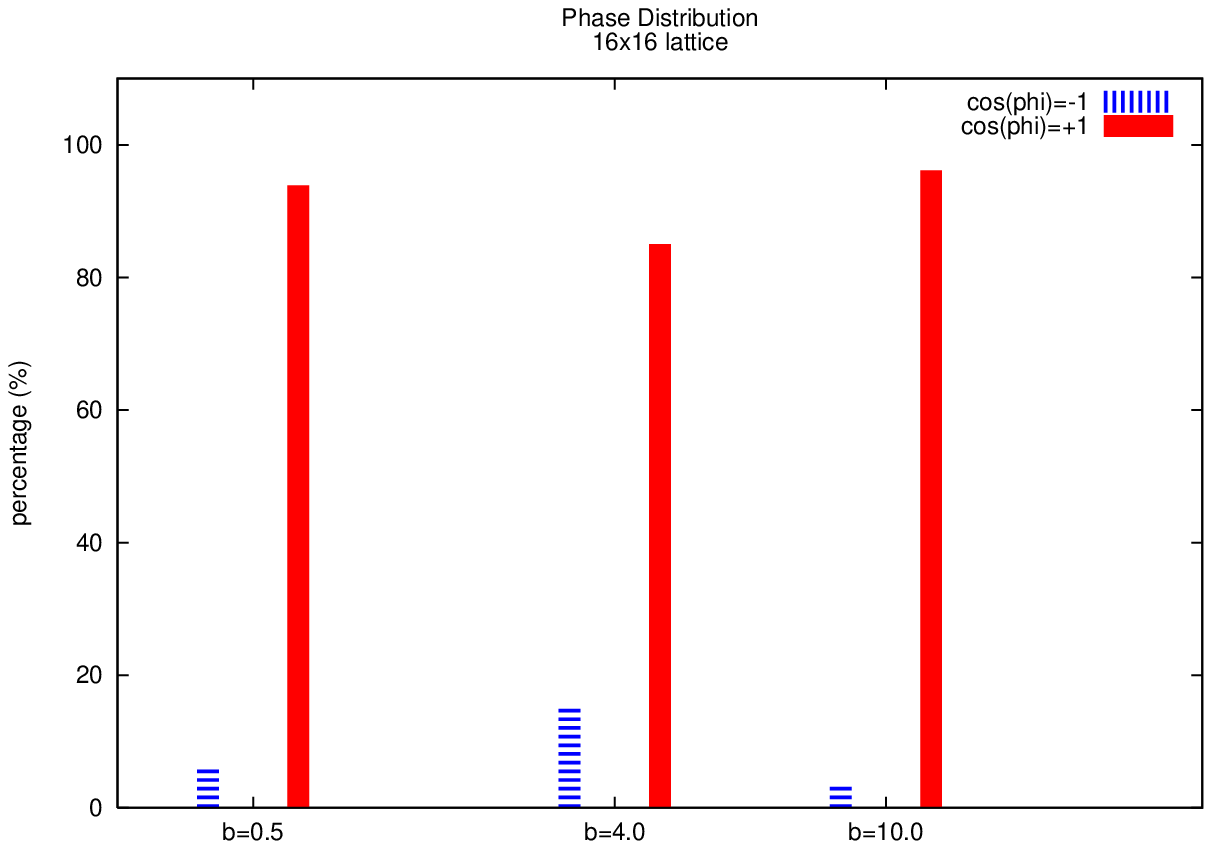}
\caption{Distribution of phase on 16x16 lattice}
\label{phasefig}
\end{figure}

Up to this point we have neglected a possible phase arising in
the calculation of fermion determinant. It is not hard to
show that the determinant is real since generically the
eigenvalues occur in complex conjugate pairs. The exception
occurs with purely real eigenvalues. As one of these crosses the
origin the sign of the determinant will change.
Indeed figure \ref{phasefig}.
indicates that this phase does undergo weak fluctuations and 
one has to check for its effect on the simulation.
As usual we can always compensate for neglecting 
the phase factor in the simulation  by re-weighting all observables
by the phase factor according to the simple rule
\begin{equation}
<O>=\frac{<Oe^{i\alpha(\Phi)}>_{\alpha=0}}{<e^{i\alpha(\Phi)>_{\alpha=0}}}
\end{equation}    
Hence we examined the 
Ward identities we have studied in the previous section now weighted 
with this phase factor. For example table~\ref{rewtable} shows the 
mean re-weighted bosonic 
action together with the phase quenched value
for different lattice sizes and a range of couplings
$\beta$. While re-weighting typically 
amplifies the estimated errors it does not appear to 
change the mean value for this observable at least within statistical errors.
This seems to indicate 
that the phase fluctuates approximately 
independently of the remaining part of the measured 
observable leading to an, at least approximate factorization in the reweighted observable
\begin{equation}
<O>_{full}=\frac{<Oe^{i\alpha}>_{\alpha=0}}{<e^{i\alpha}>_{\alpha=0}}\sim
\frac{<O>_{\alpha=0}<e^{i\alpha}>_{\alpha=0}}{<e^{i\alpha}>_{\alpha=0}}
=<O>_{\alpha=0}
\end{equation} 

\begin{table}[h]
\begin{tabular}{rl}

\begin{tabular}{|c|c|c|}
\hline
$\beta$ & $\beta<S_b>$ & $\beta<S_b>_W$\\
\hline
0.5&$145.523\pm0.286$&$144.464\pm3.858$\\
\hline
1.0&$147.369\pm0.281$&$148.764\pm21.547$\\
\hline
2.0&$135.51\pm0.281$&$139.749\pm53.707$\\
\hline
3.0&$128.275\pm0.242$&$122.731\pm8.747$\\
\hline
4.0&$127.272\pm0.329$&$126.933\pm2.093$\\
\hline
5.0&$127.326\pm0.213$&$127.342\pm1.274$\\
\hline
10.0&$127.552\pm0.299$&$127.164\pm1.967$\\
\hline
\end{tabular}

&

\begin{tabular}{|c|c|c|}
\hline
$\beta$ & $\beta<S_b>$ & $\beta<S_b>_W$  \\
\hline
0.5&$321.541\pm0.568$&$321.461\pm8.984$\\
\hline
1.0&$310.143\pm0.391$&$305.595\pm17.114$\\
\hline
2.0&$291.275\pm0.360$&$284.338\pm40.365$\\
\hline
3.0&$286.719\pm0.352$&$276.315\pm24.448$\\
\hline
4.0&$285.546\pm0.326$&$283.904\pm5.489$\\
\hline
5.0&$285.627\pm0.340$&$285.307\pm3.386$\\
\hline
10.0&$287.163\pm0.285$&$286.943\pm2.483$\\
\hline
\end{tabular}
\\
\\
\multicolumn{2}{c}{
\begin{tabular}{|c|c|c|}
\hline
$\beta$ & $\beta<S_b>$ & $\beta<S_b>_W$  \\
\hline
0.5&$283.02\pm2.166$&$285.369\pm4.369$\\
\hline
1.0&$458.823\pm1.707$&$459.028\pm7.847$\\
\hline
2.0&$509.266\pm0.628$&$504.264\pm38.233$\\
\hline
3.0&$503.627\pm0.784$&$496.215\pm46.938$\\
\hline
4.0&$507.243\pm0.609$&$505.991\pm13.837$\\
\hline
5.0&$507.412\pm1.028$&$506.915\pm14.778$\\
\hline
10.0&$511.561\pm0.570$&$511.715\pm5.386$\\
\hline
\end{tabular}
}

\end{tabular}
\caption{$\beta<S_b>$ for quenched and unquenched phase on 8x8, 12x12 and 16x16 lattices.}
\label{rewtable}
\end{table}

\subsection{Continuum Limit}
\begin{figure}
\centering
\includegraphics[width=8cm,angle=-90]{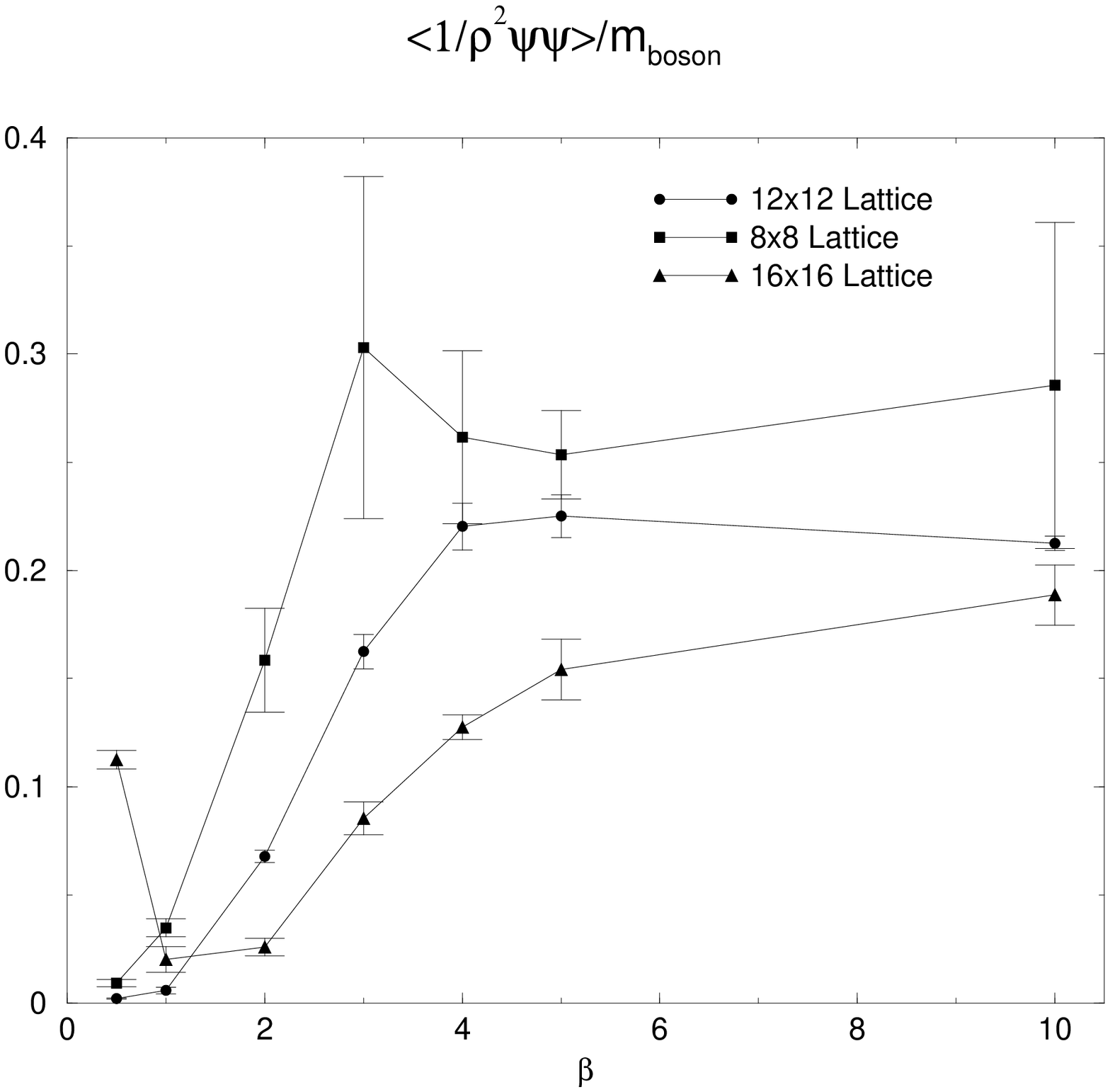}
\caption{$<1/\rho^2\overline{\psi}\psi>/m^B$}
\label{condbymfig}
\end{figure}

Figure \ref{condbymfig} shows a plot for the dimensionless ratio
value 
$\frac{<1/\rho^2\overline{\psi}\psi>}{m_B}$ ($m_B$ is the lightest
boson mass)
for different lattice sizes and for different values of the coupling. 
One can see that an approximate plateau occurs for sufficiently
small lattice spacing which we interpret as the onset of continuum
physics. The separation of the curves from different lattice volumes
we consider to be a measure of possible finite volume effects. 
Modulo these finite volume questions we take this as evidence that
the magnitude of the condensate is non-zero in the continuum
limit.

\section{Conclusions}
This paper is devoted to a study of the ${\cal N}=2$ supersymmetric
sigma model regulated on a lattice. The lattice formulation
we employ was derived in an earlier paper \cite{sigma1} 
and results from a discretization of a {\it twisted} version 
of the continuum theory. The twisting process is to be viewed
as a change of variables (we are in flat space) and has the merit
of exposing a nilpotent scalar supersymmetry $Q$ which can be
preserved under discretization. In this paper we extend our previous
work by introducing a modified Wilson operator in the form of
a twisted mass term which is used to remove doubler modes from
our lattice theory without spoiling the $Q$-exact property of
the lattice action. The introduction of such a mass term is
compatible with supersymmetry if it takes the form of a 
holomorphic Killing vector in the target space \cite{freedman}
and corresponds to the addition of central charges in the
supersymmetry algebra. The replacement of a simple mass term with
a Wilson difference operator breaks supersymmetry softly and
is not expected to lead to additional fine tuning. 
We study the $CP^1\sim O(3)$-model 
in detail using the RHMC algorithm on lattices
as large as $16^2$ over a range of coupling $\beta=0.5\to 10.0$.
Our results for both the spectrum and
Ward identities provide strong evidence for a restoration of {\it full}
supersymmetry at large $\beta$ and small lattice spacing {\it without
additional fine tuning}. Notice that while the Ward identities 
are examples of trivial topological observables the mass spectrum
we measure is non-topological. Further investigations of such physical
observables are underway and will published elsewhere \cite{future}.
We are also able to see a non-zero chiral condensate 
and this appears to persist into the continuum limit as expected from
the chiral anomaly. 
It would be very interesting to extend this work to the general
$CP^{N-1}$ models where the numerical results could be compared
with exact results on the low lying mass spectrum \cite{hollowood}.

\section{Acknowledgements}
This work was supported in part by DOE grant DE-FG02-85ER40237.

\newpage
\appendix
\section{Appendix: Notations}
\label{nota}

A real system is denoted by $x^1$, $x^2$ which combine to give holomorphic coordinates,
\begin{equation} 
z=x^1+ix^2, \quad \bar{z}=x^1-ix^2 
\end{equation} 
Then a vector $V^\mu$ with components $(V^1,V^2)$ has holomorphic components,
\begin{equation}
V_+=\frac{1}{2}(V_1-iV_2), \quad V_-=\frac{1}{2}(V_1+iV_2) 
\end{equation}  
The Dirac matrices $\gamma^\mu$ are given by (Chiral basis),
\begin{equation}
(\gamma^1)_{\alpha}^{\;\;\beta}=\sigma^1,\quad (\gamma^2)_{\alpha}^{\;\;\beta}=\sigma^2
\end{equation}
Spinor indices are raised and lowered by the matrix $C_{\alpha\beta}=\sigma^1$,
\begin{equation}
(\gamma^\mu)_{\alpha\beta}=(\gamma^\mu)_\alpha^{\;\;\tau}C_{\tau\beta}
\end{equation}
thus for a vector $V^\mu$,
\begin{eqnarray}
\gamma^\mu_{++} V_\mu = V_1 - i V_2 = V_+ \\
\gamma^\mu_{--} V_\mu = V_1 + i V_2 = V_- \\
\end{eqnarray} 
For a K\"ahler manifold, the metric is given by,
\begin{equation}
g_{I\bar{J}}=\partial_I\partial_{\bar{J}}K 
\end{equation}
where $K$ is a K\"ahler potential.
(Note that $g_{\bar{I}\bar{J}}=g_{IJ}=0$ for a almost complex manifold).\\
The non vanishing components of the Christoffel symbols are then,
\begin{equation}
\Gamma^I_{JK}=g^{I\bar{L}}\partial_Jg_{K\bar{L}}, \quad 
\Gamma^{\bar{I}}_{\bar{J}\bar{K}}=g^{\bar{I}L}\partial_{\bar{J}}g_{L\bar{K}}
\end{equation}
This implies the only non-trivial components of the Riemann tensor are,
\begin{equation}
R_{\bar{I}J\bar{K}L}=g_{\bar{I}M}\partial_{\bar{K}}\Gamma^M_{JL},\; or \quad  
R^I_{\;\;J\bar{K}L}=\partial_{\bar{K}}\Gamma^I_{JL},
\end{equation}

\section{Appendix: Twisted ${\cal N}=2$ supersymmetry algebra}
\label{twisted}
The algebra of ${\cal N}=2$ supersymmetry in 2 space
dimensions is given by \cite{labast},
\begin{equation}
\begin{array}{lll}
\{Q_{\alpha+},Q_{\beta-}\}=\gamma^\mu_{\alpha\beta}P_\mu, &\quad& [ R,Q_{\alpha\pm} ]=
\pm\frac{1}{2}Q_{\alpha\pm} \\
\{Q_{\alpha+},Q_{\beta+}\}=\{Q_{\alpha-},Q_{\beta-}\}=0,  &\quad& [ J,P_\mu ] = 
-i\epsilon_\mu^{\;\nu}P_{\nu} \\
\lbrack Q_{\alpha a},P_\mu ]=[P_\mu,P_\nu]=0,             &\quad& [R,P_\mu]=0 \\
\lbrack J,Q_{\pm a}]=\pm\frac{1}{2}Q_{\pm a},              &\quad& [J,R]=[J,J]=[R,R]=0 
\end{array}
\label{orig}  
\end{equation}
where J is the generator of Lorentz $SO(2)$ symmetry and R is the generator of the internal
$SO(2)$ symmetry. In order to obtain some scaler charges, one perform a twisting that 
changes the spin of the above charges. In order to do this we need to redefine the Lorentz 
generator. Indeed if we take (see Appendix~\ref{nota} for our notations),
\begin{equation}
\tilde{J}=J+R
\end{equation}
with respect to this new generator one finds that $Q_{+-}$ and $Q_{-+}$ behave as scalars 
while the pair $Q_{++}$ and $Q_{--}$ as vectors. To make manifest the new Lorentz 
structure of each of the generators we define,
\begin{displaymath}
\begin{array}{l}
Q_L = Q_{+-} \\
Q_R = Q_{-+} \\
Q_{++}= \gamma^\mu_{++}G_\mu = G_+\\
Q_{--}= \gamma^\mu_{--}G_\mu = G_- \\
\end{array}
\end{displaymath}
it is clear from (\ref{orig}) that,
\begin{equation}
Q_L^2=Q_R^2=\{Q_L,Q_R\}=0
\end{equation}
The algebra (\ref{orig}) in terms of the new Lorentz generator $\tilde{J}$ and the following 
redefined operators,
\begin{equation}
Q=Q_L+Q_R,\quad M=Q_L-Q_R
\end{equation}
is
\begin{equation}
\label{twistAlg}
\begin{array}{l}
Q^2=M^2=\{Q,M\}=[Q,P_\mu]=[M,P_\mu]=0 \\
\{Q,G_\mu\}=P_\mu \\
\lbrack Q,\tilde{J}]=[M,\tilde{J}]=0 \\
\{M,G_\mu\}=-i\epsilon_\mu^{\;\;\nu}P_\nu \\
\lbrack \tilde{J},P_\mu]=-i\epsilon_\mu^{\;\;\nu}P_\nu \\
\lbrack \tilde{J},G_\mu]=-i\epsilon_\mu^{\;\;\nu}G_\nu \\
\lbrack P_\mu,P_\nu]=\{G_\mu,G_\nu\}=[\tilde{J},\tilde{J}]=0 
\end{array}
\label{twist}
\end{equation}
in addition, the action of the R generator on the new twisted charges is,
\begin{equation}
\begin{array}{l}
\lbrack R,Q]=-\frac{1}{2} M\\
\lbrack R,M]=-\frac{1}{2} Q\\
\lbrack R,G_\mu]=-\frac{i}{2} \epsilon_\mu^{\;\;\nu} G_\nu\\
\lbrack R,R]=[R,\tilde{J}]=[R,P_\mu]=0\\
\end{array}
\end{equation}

Let our field content be $\phi^i, B^i_\alpha, \psi^i$ and $\eta^i_\alpha$ or on the 
complex manifold $\phi^I, B^I_+, \psi^I$ and $\eta^I_+$. The R-transformations of these 
fields is given by
\begin{equation}
\begin{array}{lll}
\lbrack R,\phi^I]=0,     &\quad& [R,\phi^{\bar{I}}]=0\\
\lbrack R,\psi^I]=\frac{1}{2} \psi^I, &\quad& [R,\psi^{\bar{I}}]=-\frac{1}{2}
\psi^{\bar{I}}\\
\lbrack R,\eta^I_+]=\frac{1}{2} \eta^I_+, &\quad& [R,\eta^{\bar{I}}_-]=
-\frac{1}{2} \eta^{\bar{I}}_-\\
\lbrack R,B^I_+]=B^I_+,   &\quad& [R,B^{\bar{I}}_-]=-B^{\bar{I}}_-\\
\end{array}
\end{equation}

More importantly the transformation of our fields under the twisted charges $Q,M,G_+$ and 
$G_-$
is given by, for $Q$, 
\begin{equation}
\begin{array}{lll}
Q\phi^I = \psi^I &\quad& Q\phi^{\bar{I}} = \psi^{\bar{I}}\\
Q\psi^I = 0 &\quad& Q\psi^{\bar{I}}= 0\\
Q\eta^I_+=B^I_+-\Gamma^I_{KJ}\psi^J\eta^K_+ &\quad& 
    Q\eta^{\bar{I}}_-=B^{\bar{I}}_--\Gamma^{\bar{I}}_{\bar{K}\bar{J}}
\psi^{\bar{J}}\eta^{\bar{K}}_-\\
QB^I_+ = -\Gamma^I_{JK}\psi^JB^K_+-R^I_{\;\;K\bar{J}L}\psi^K\psi^{\bar{J}}\eta^L_+ &\quad& 
    QB^{\bar{I}}_- = -\Gamma^{\bar{I}}_{\bar{J}\bar{K}}\psi^{\bar{J}}
    B^{\bar{K}}_--R^{\bar{I}}_{\;\;\bar{J}L\bar{K}}\psi^{\bar{J}}\psi^L
\eta^{\bar{K}}_-\\
\end{array}
\end{equation}
for $M$
\begin{equation}
\begin{array}{lll}
M\phi^I = -\psi^I &\quad& M\phi^{\bar{I}} = \psi^{\bar{I}}\\
M\psi^I = 0 &\quad& M\psi^{\bar{I}}= 0\\
M\eta^I_+=-(B^I_+-\Gamma^I_{KJ}\psi^J\eta^K_+) &\quad& 
    M\eta^{\bar{I}}_-=B^{\bar{I}}_--\Gamma^{\bar{I}}_{\bar{K}\bar{J}}
\psi^{\bar{J}}\eta^{\bar{K}}_-\\
MB^I_+ = \Gamma^I_{JK}\psi^JB^K_+-R^I_{\;\;K\bar{L}J}\psi^J\psi^{\bar{L}}\eta^K_+ &\quad& 
    MB^{\bar{I}}_- = -\Gamma^{\bar{I}}_{\bar{J}\bar{K}}\psi^{\bar{J}}
    B^{\bar{K}}_-+R^{\bar{I}}_{\;\;\bar{K}L\bar{J}}\psi^{\bar{J}}\psi^L
\eta^{\bar{K}}_-\\
\end{array}
\end{equation}
for $G_+$
\begin{equation}
\begin{array}{lll}
G_+\phi^I = \frac{1}{2}\eta^I_+ &\quad& G_+\phi^{\bar{I}} = 0\\
G_+\psi^I = -\frac{1}{2}(B^I_+-\Gamma^I_{KJ}\psi^J\eta^K_+) &\quad& G_+\psi^{\bar{I}}= 0\\
G_+\eta^I_+=0 &\quad& G_+\eta^{\bar{I}}_-=0\\
G_+B^I_+ = -\frac{1}{2}\Gamma^I_{JK}B^J_+ \eta^k_+&\quad& G_+B^{\bar{I}}_- = 
\partial_+\eta^{\bar{I}}_--\frac{1}{2}R^{\bar{I}}_{\;\;\bar{K}L\bar{J}}
\psi^{\bar{J}}\eta^L_+\eta^{\bar{K}}_-\\
\end{array}
\end{equation}
and finally for $G_-$,
\begin{equation}
\begin{array}{lll}
G_-\phi^I = 0 &\quad& G_-\phi^{\bar{I}} = \frac{1}{2}
\eta^{\bar{I}}_-\\
G_-\psi^I = 0 &\quad& G_-\psi^{\bar{I}}= -\frac{1}{2}(B^{\bar{I}}_-
                    -\Gamma^{\bar{I}}_{\bar{K}\bar{J}}\psi^{\bar{J}}
\eta^{\bar{K}}_-)\\
G_-\eta^I_+=0 &\quad& G_-\eta^{\bar{I}}_-=0\\
G_-B^I_+ = \partial_-\eta^I_+-\frac{1}{2}R^I_{\;\;K\bar{L}J}
\psi^J\eta^{\bar{L}}_-\eta^K_+&\quad& 
G_-B^{\bar{I}}_- = -\frac{1}{2}\Gamma^{\bar{I}}_{\bar{J}\bar{K}}
B^{\bar{J}}_- \eta^{\bar{k}}_- 
\\
\end{array}
\end{equation}

%%%%%%%%%%%%%%%%%%%%%%%%%%%%%%%%%%%%%%%%%%%%%%%%%%%%%%%%%%%%%%%%%%%%5) 
% figures                                                          %
%%%%%%%%%%%%%%%%%%%%%%%%%%%%%%%%%%%%%%%%%%%%%%%%%%%%%%%%%%%%%%%%%%%%

%spectrum
%%%%%%%%%

%\beta<S_B>/(2L^2)
%%%%%%%%%%%%%%%%%

%<QO>=0
%%%%%%%

%<GO>=0
%%%%%%%

% old WI Q \eta_x \bar{u}_y
%%%%%%%%%%%%%%%%%%%%%%%%%%

% condensate
%%%%%%%%%%%%

% phase
%%%%%%%

% cond by m
%%%%%%%%%%%

\end{document}